\documentclass[twocolumn]{aastex63}

\newcommand{\thi}{$\tau_{\rm HI}(v)$}
\newcommand{\hi}{H\textsc{i}}
\newcommand{\nhi}{$N(\rm H\textsc{i})$}
\newcommand{\nhithin}{$N(\rm H\textsc{i})^*$}
\newcommand{\rhi}{$\mathcal{R}_{\rm HI}$}
\newcommand{\fcnm}{$f_{\rm CNM}$}

\newcommand{\R}{$\mathcal{R}$}
\newcommand{\nabs}{$44$}
\newcommand{\noff}{$40$}

\newcommand{\nrest}{$103$}

\submitjournal{ApJS}

\shorttitle{MACH}
\shortauthors{Murray et al.}
\graphicspath{{./}{figures/}}
\begin{document}

\title{The MACH HI absorption survey I: Physical conditions of cold atomic gas outside of the Galactic plane}

\correspondingauthor{C.\,E.\,M.}
\email{clairemurray56@gmail.com}

\author[0000-0002-7743-8129]{Claire E. Murray}
\altaffiliation{NSF Astronomy \& Astrophysics Postdoctoral Fellow}
\affil{Department of Physics \& Astronomy, 
Johns Hopkins University,
3400 N. Charles Street, 
Baltimore, MD 21218}

\author[0000-0002-3418-7817]{Sne\v{z}ana Stanimirovi\'c}
\affil{Department of Astronomy, 
University of Wisconsin -- Madison, 
475 N. Charter Street,
Madison, WI 53706}

\author[0000-0002-7456-8067]{Carl Heiles}
\affil{Department of Astronomy, University of California, Berkeley, 601 Campbell Hall 3411, Berkeley,
CA 94720}

\author[0000-0002-6300-7459]{John M. Dickey}
\affil{School of Natural Sciences, University of Tasmania, Hobart, TAS 7001, Australia}

\author[0000-0003-2730-957X]{N.\,M. McClure-Griffiths}
\affil{Research School of Astronomy and Astrophysics - The Australian National University, Canberra, ACT, 2611, Australia}

\author[0000-0002-9888-0784]{M.-Y. Lee}
\affil{Korea Astronomy and Space Science Institute, 776 Daedeokdae-ro, 34055 Daejeon, Republic of Korea}

\author{W.\, M. Goss}
\affil{National Radio Astronomy Observatory, P.O. Box O, 1003 Lopezville, Socorro, NM 87801, USA}

\author{Nicholas Killerby-Smith}
\affil{Research School of Astronomy and Astrophysics - The Australian National University, Canberra, ACT, 2611, Australia}

\begin{abstract}
Tracing the transition between the diffuse atomic interstellar medium (ISM) and cold, dense gas is crucial for deciphering the star formation cycle in galaxies. Here we present MACH, a new survey of cold neutral hydrogen (\hi) absorption at $21\rm\,cm$ by the Karl G. Jansky Very Large Array. We target 42 bright background sources with $60<l<110^{\circ}$, $30<b<62^{\circ}$, significantly expanding the sample of publicly-available, sensitive $21\rm\,cm$ absorption outside the Galactic plane. With matching $21\rm\,cm$ emission data from the EBHIS survey, we measure the total column density and cold \hi\ fraction, and quantify the properties of individual \hi\ structures along each sightline via autonomous Gaussian decomposition. Combining the MACH sample with results from recent \hi\ absorption surveys, we produce a robust characterisation of the cool atomic medium at high and intermediate Galactic latitudes. We find that MACH \hi\ has significantly smaller column density relative to samples at similar latitudes, and the detected cold \hi\ structures have smaller line widths, temperatures and turbulent Mach numbers, suggesting that MACH probes a particularly quiescent region. Using all available observations, we compute the cumulative covering fraction ($c$) of cold \hi\ at local velocities outside the disk: structures with $\tau>0.001$ are ubiquitous ($c\sim100\%$), whereas high optical depths ($\tau>1$) are extremely rare ($c\sim0\%$). 
\end{abstract}


\section{Introduction} 
\label{sec:intro}

As stars form and evolve in the interstellar medium (ISM), they generate a rich, multi-phase structure of gas and dust via radiative and dynamical feedback. To follow the mass flow between gas reservoirs and star formation in galaxies, it is essential to resolve the nature and influence of these feedback mechanisms \citep{hopkins2014, gatto2017}

The properties of neutral hydrogen (\hi), essential fuel for the evolution of star-forming clouds, bear clues to the effects of feedback in the ISM. From a theoretical perspective, we expect \hi\ to occupy multiple phases, including the cold ($T\sim20-200\rm\,K$) and warm ($T\sim1000-8000\rm\,K$) neutral media \citep[CNM and WNM;][]{mckee1977, wolfire2003}. However, the mass distribution of \hi\ between these phases depends strongly on the nature of turbulent, radiative, and dynamical processes in the ISM. For example, the density and temperature of atomic gas surrounding molecular clouds strongly depends on the initial speed of colliding gas flows \citep[e.g.,][]{ntormousi2011, clark2012}, and/or whether supernovae explode within density peaks or randomly \citep[e.g.,][]{gatto2015}. Beyond the disk, the CNM content of accretion streams and outflows in galactic halos is determined by star formation feedback launching galactic winds \citep[e.g., ][]{fauchergiguere2015}. 

Detailed measurements of CNM and WNM properties are useful for distinguishing between these disparate theoretical pictures of the ISM. However, observational constraints require measurements of \emph{both} emission and absorption. The first observations of \hi\ absorption at $21\rm\,cm$ confirmed that \hi\ is organized into distinct phases: the CNM which absorbs strongly, and the pervasive WNM which does not \citep{clark1965, dickey1978, heiles1980}. The spectral line widths of cool, absorbing clouds (CNM) are narrower than those observed in emission \citep{mebold1972, radhakrishnan1972, dickey1978, liszt1983, roy2013, murray2015} further emphasizing temperature variations between \hi\ structures. Furthermore, the properties of the CNM appear to be uniform and do not vary significantly between diffuse, high-latitude regions and the Galactic plane \citep[e.g.,][]{dickey1981}. However, the fraction of CNM increases toward dense, molecular cloud environments and with the total \hi\ column density \citep{stanimirovic2014, nguyen2019}. 

\begin{figure*}
\begin{center}
\vspace{-60pt}
\includegraphics[width=\textwidth]{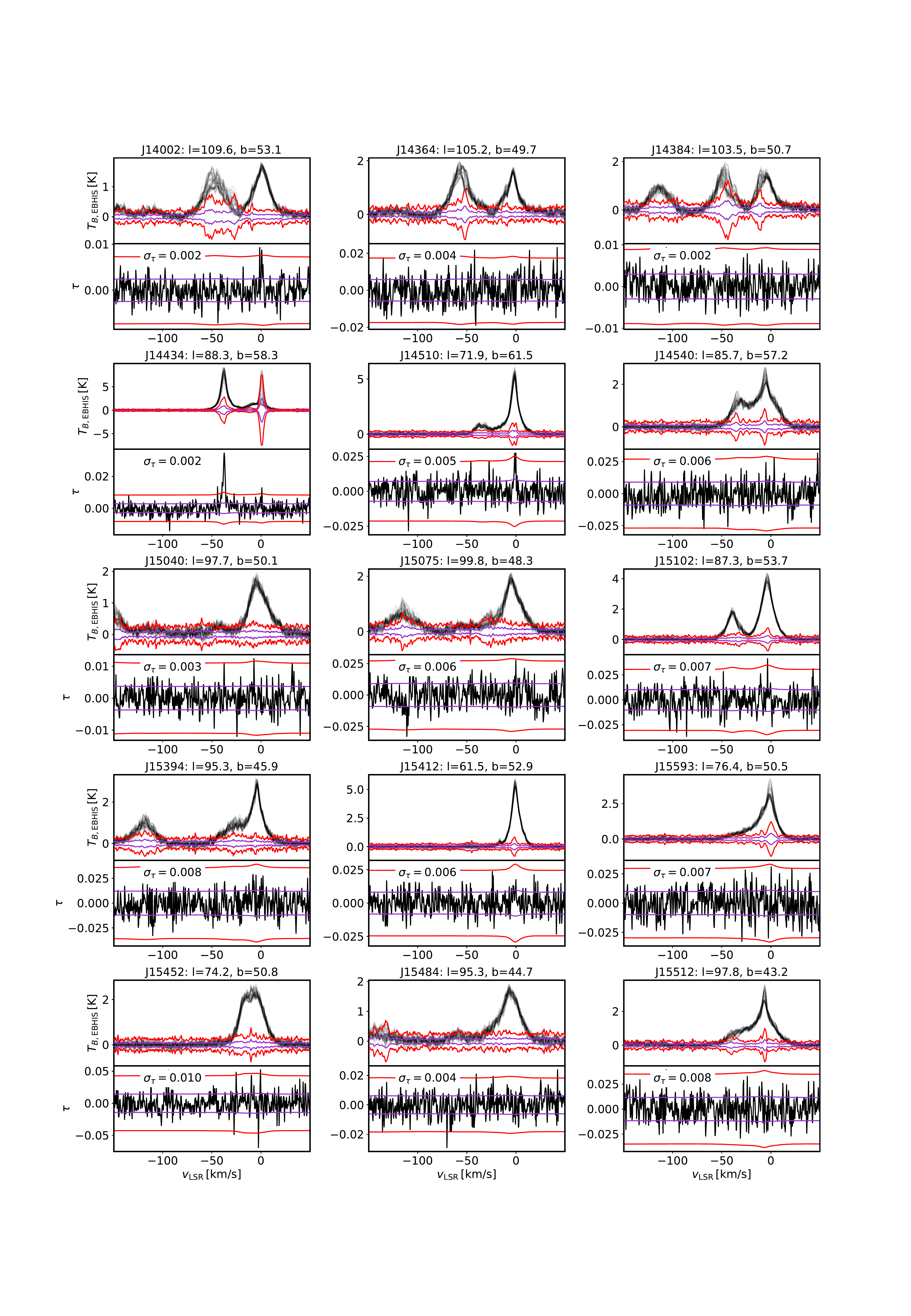}
% \vspace{-70pt}
\caption{Each panel includes the $21\rm\,cm$ emission ($T_B(v)$; upper) from EBHIS \citep{winkel2016}, and $21\rm\,cm$ optical depth (\thi; lower) for the \nabs\ targets in the MACH survey. The median uncertainties in optical depth are printed within each lower panel, and the uncertainties as a function of velocity for $T_B(v)$ and \thi\ are overlaid in both panels ($\pm1\sigma$; purple;  $\pm3\sigma$, red). }
\label{f:spectra_summary}
\end{center}
\end{figure*}

\begin{figure*}
% \ContinuedFloat
\begin{center}
\vspace{-60pt}
\includegraphics[width=\textwidth]{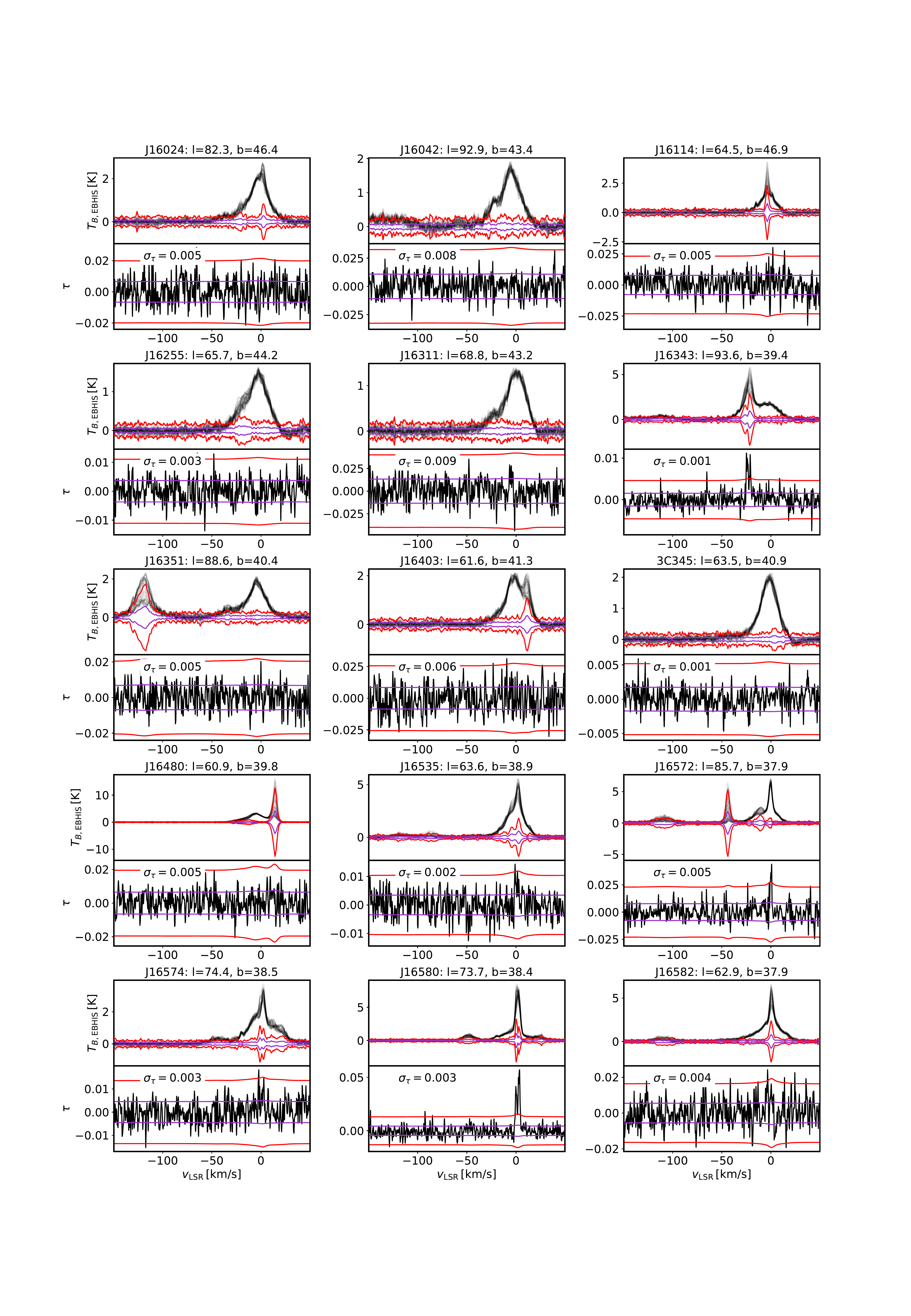}
% \vspace{-70pt}
% \caption{(contd.) Each panel includes the $21\rm\,cm$ emission ($T_B(v)$; upper) from EBHIS \citep{winkel2016}, and $21\rm\,cm$ optical depth (\thi; lower) for the \nabs\ targets in the MACH survey. The median uncertainties in optical depth are printed within each lower panel, and the uncertainties as a function of velocity for $T_B(v)$ and \thi\ are overlaid in both panels ($\pm1\sigma$; purple;  $\pm3\sigma$, red). }
% % \label{f:spectra_summary_b}
\end{center}
\end{figure*}

\begin{figure*}
% \ContinuedFloat
\begin{center}
\vspace{-60pt}
\includegraphics[width=\textwidth]{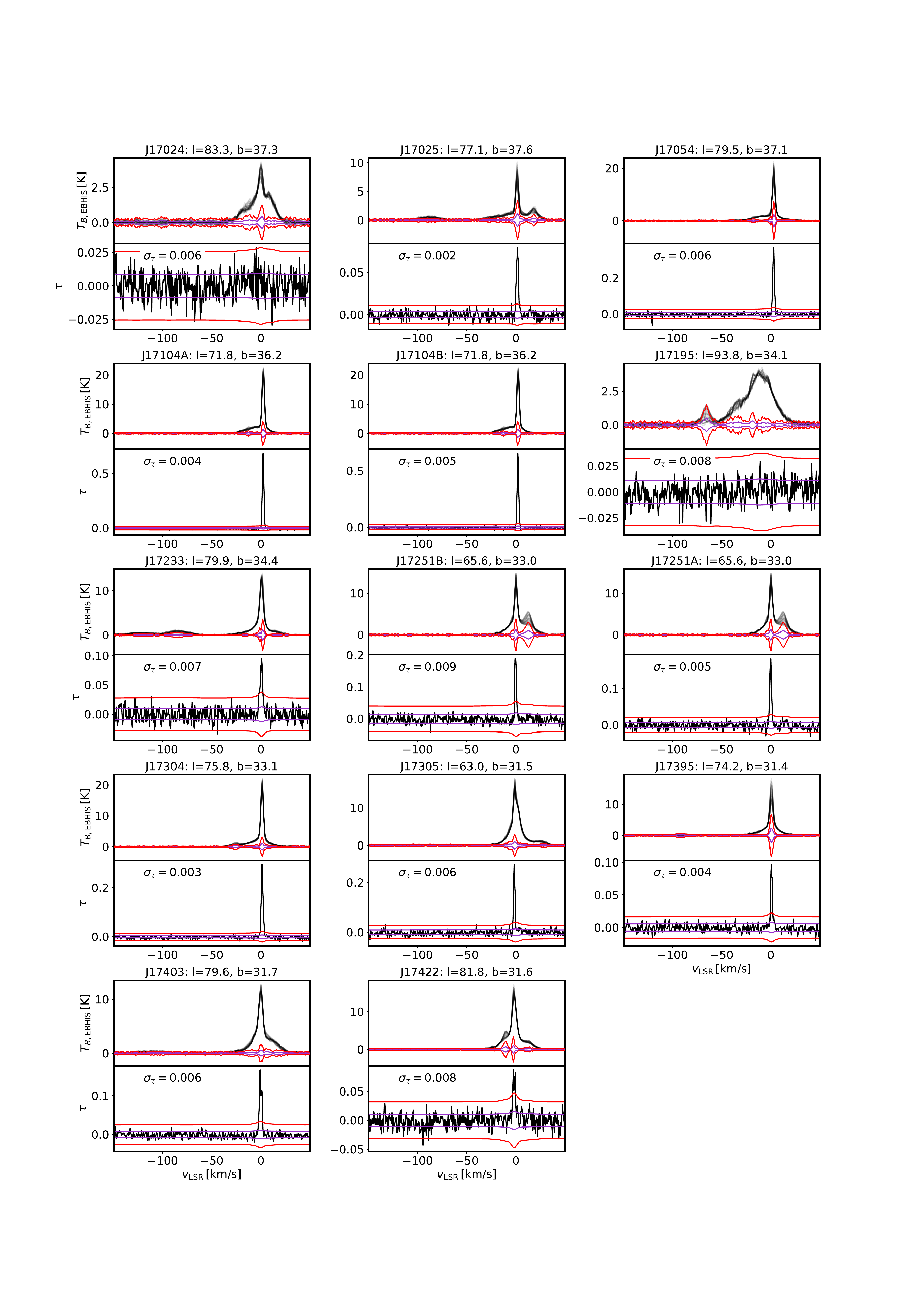}
% \vspace{-70pt}
% \caption{(contd.) Each panel includes $21\rm\,cm$ emission ($T_B(v)$; upper) from EBHIS \citep{winkel2016}, and $21\rm\,cm$ optical depth (\thi; lower) for the \nabs\ MACH LOS. The median uncertainties in optical depth are printed within each lower panel, and the uncertainties as a function of velocity for $T_B(v)$ and \thi\ are overlaid in both panels ($\pm1\sigma$; purple;  $\pm3\sigma$, red). }
% \label{f:spectra_summary}
\end{center}
\end{figure*}

Despite the advances of $21\rm\,cm$ absorption studies to date, they are limited by the availability of the brightest background continuum sources for measuring absorption with sufficient sensitivity, and are therefore sparsely distributed. In addition, given the ubiquity of \hi\ in the Milky Way, the significant blending of $21\rm\,cm$ spectra can make it nearly impossible to distinguish individual \hi\ structures. Detailed comparisons between real $21\rm\,cm$ observations and synthetic observations of numerical simulations have shown that the completeness of ``cloud" recovery declines severely with decreasing Galactic latitude due to crowding of \hi\ spectral line profiles in velocity \citep{murray2017}. At high latitudes the line of sight path length through the global cool \hi\ layer is just a few times the CNM scale height ($h\sim120\rm\, pc$), providing our best opportunity to reliably measure the properties of individual \hi\ structures. These properties include the column density, temperature and turbulent Mach number, all of which provide important benchmarks for numerical models of the ISM \citep[e.g.,][]{villagran2018}.

In addition, constraining the properties of \hi\ at high latitude is important for sorting out the ISM mass budget. It is clear that \hi\ and carbon monoxide (CO) emission observations are missing significant quantities of gas in galaxies traced by dust and gamma ray emission \citep[e.g., ][]{grenier2005, planck2011}. This so-called ``CO-dark” gas can be accounted for in several ways, including by poorly-shielded H$_2$ molecules, variations in dust grain emissivity (which controls the conversion between dust emission and mass), or optically-thick \hi\ \citep[e.g.,][]{reach2017a, reach2017b}. Although data from previous \hi\ absorption studies is sufficient to statistically rule out the hypothesis that optically-thick \hi\ dominates dark gas \citep{murray2018a}, quantifying its influence requires building expanded samples which probe diverse Galactic environments.

\begin{figure}
\begin{center}
\includegraphics[width=0.53\textwidth]{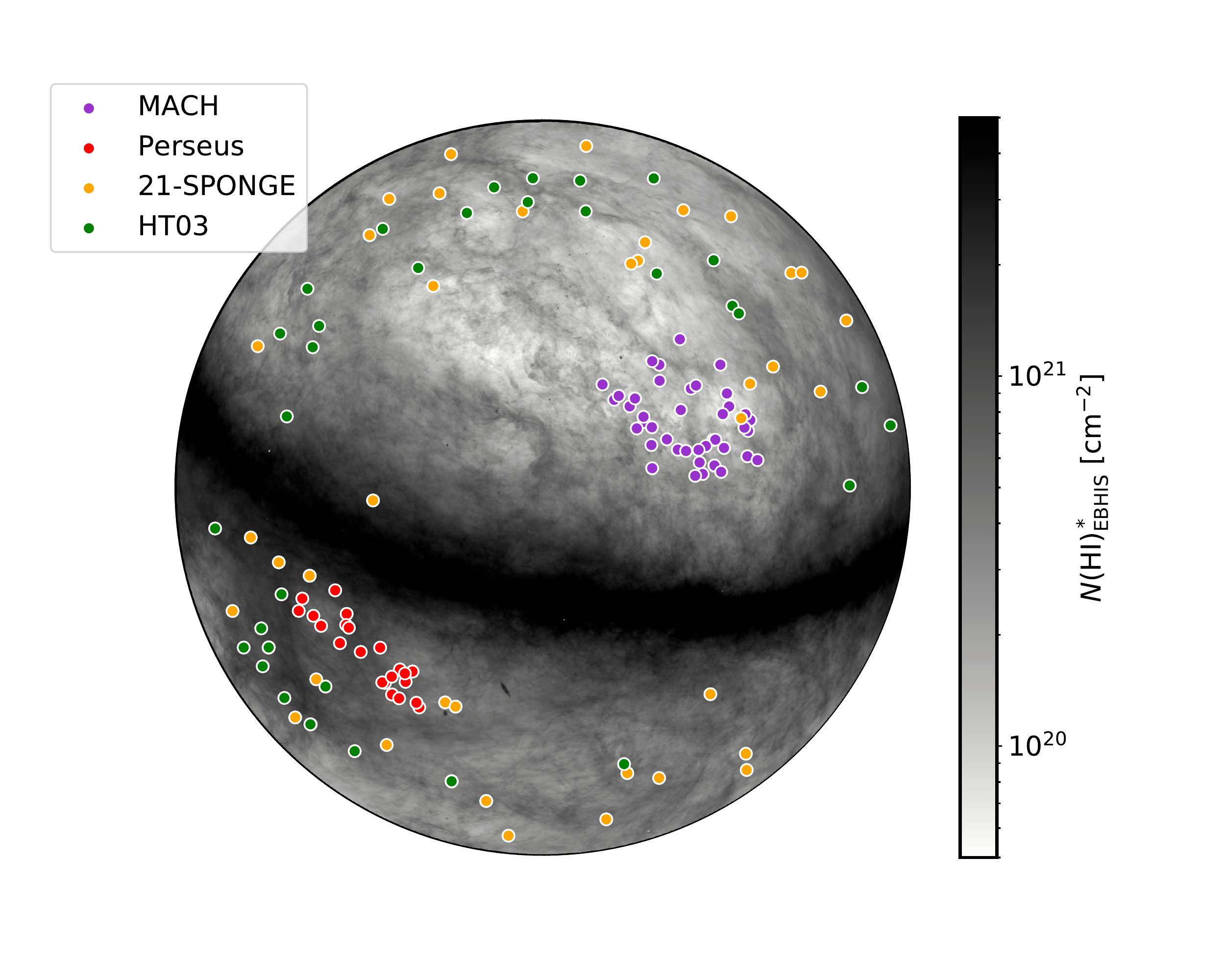}
\caption{Distribution of MACH target sources (purple) overlaid on a zenith-equal-area (ZEA) map of the \hi\ column density from EBHIS \citep{winkel2016}. The target sources comprising the comparison sample from \citet[][``Perseus"]{stanimirovic2014}, \citet[][``21-SPONGE"]{murray2018b} and \citet[][HT03]{heiles2003} are included (red, green and orange circles, respectively). }
\label{f:position_comparison}
\end{center}
\end{figure}

In this work, we present data from Measuring Absorption by Cold Hydrogen (MACH) -- a survey of $21\rm\,cm$ absorption at high Galactic latitude with the Karl G. Jansky Very Large Array (VLA). MACH increases the sample of publicly available, high-sensitivity $21\rm\,cm$ absorption spectra by $50\%$. The paper is organized as follows. In Section~\ref{sec:data} we discuss the VLA observations and data reduction strategy, including extraction of matching $21\rm\,cm$ emission. In Section~\ref{sec:analysis}, we present our analysis methodology, including computing integrated \hi\ properties along MACH lines of sight (LOS) and decomposing MACH LOS into individual \hi\ structures via autonomous Gaussian decomposition. In Section~\ref{sec:results}, we present the results of our analysis, including a parallel, identical analysis of available high-latitude $21\rm\,cm$ absorption from the literature.  In Section~\ref{sec:discussion}, we discuss the comparison between MACH and the rest of the high-latitude sky, and estimate the covering fraction of cold \hi. Finally, in Section~\ref{sec:summary} we summarize the results and present our conclusions. 

\section{Data}
\label{sec:data}

\subsection{Observations}

For the MACH survey, we targeted extragalactic continuum sources in the region defined by $30<b<62^{\circ}$, $60<l<110^{\circ}$. This region was selected to probe absorption by the local ISM at high latitude, and to overlap with the high-velocity cloud Complex C.\footnote{We note that the observing setups for all targets did not cover the complete velocity range of Complex C. As a result, we will defer the analysis of the high-velocity data to a future paper once complete coverage is obtained.}  We selected the $42$ sources from the VLA-FIRST survey \citep{becker1995} with bright flux density at $21\rm\,cm$ ($S_{\rm 21cm}>0.45 \rm\,Jy$) and source size estimates $<5^{\prime \prime}\times5^{\prime \prime}$. These parameters were chosen to maximize sensitivity to absorption and ensure that the majority of our targets would be unresolved.

Observations were conducted at the VLA between August 2017 and February 2018 and spanned several VLA configurations, including C, B, and BnA, as well as configuration moves (C to B, BnA to A). We observed 42 target sources in 51 hours, for an average of $\sim50$ minutes per target. Each observation utilized three separate, standard L-band configurations, each with one dual-polarization intermediate frequency band of width $1\rm\,MHz$ and $1.95\rm\, kHz$ per channel spacing. The target band was centered on the \hi\ line ($1.42040575\rm\,GHz$) at a velocity in the local standard of rest (LSR) of $v_{\rm LSR}=-50\rm\,km\,s^{-1}$ and two offline bands were centered at $\pm1 \rm\,MHz$ respectively. The offline bands were used to perform bandpass calibration via frequency switching, as \hi\ absorption at Galactic velocities in the direction of our calibrator sources can be significant \citep{murray2015, murray2018b}. The absolute phase change associated with frequency switching does not affect our results, as we normalized our solutions with respect to the continuum (see below). Our setup resulted in a velocity coverage of $200\rm\,km\,s^{-1}$ ($-150\rm\,km\,s^{-1}-50\rm\,km\,s^{-1}$) with $0.42\rm\,km\,s^{-1}$ channel spacing, which corresponds to $\sim0.5\rm\,km\,s^{-1}$ velocity resolution \citep{rohlfs2004}. We observed nearby VLA calibrator sources for phase and amplitude calibration, and employed self-calibration on each target source for relative flux calibration.

\subsection{Data Reduction}

Following the strategy of the 21-SPONGE survey, we reduced all MACH data using the Astronomical Image Processing System \citep[AIPS;][]{greisen2003}. As a first step, all baselines shorter than $300\rm\,m$ were excluded to avoid contamination from partially-resolved \hi\ emission. After interactive flagging of noisy baselines, time intervals and antennas using AIPS task TVFLG, we computed an initial bandpass calibration solution for each of the frequency-switched subbands using AIPS task BPASS, and examined them with AIPS task BPLOT to verify good solutions for each antenna. We then combined the offline subbands to create a final bandpass solution to apply to the target. 

Turning to the target data set, we performed amplitude and phase calibration with AIPS task CALIB. Next, we determined the relative flux calibration using self-calibration on the target source. This involves isolating the target source continuum, constructing an image using AIPS task IMAGR and using it to calibrate the continuum data set. This process was repeated until the signal to noise in the continuum image following each round of calibration no longer improved significantly (typically requires $1-2$ iterations). Next, we subtracted the continuum from the target by fitting a linear model to line-free channels with AIPS task UVLSF, corrected the target source velocities for Earth's rotation and LSR motion using AIPS task CVEL, and finally applied the self-calibration solution to the target. 

Following flagging and calibration, we constructed final image cubes using IMAGR. We estimated the background noise for each target using a test image of one channel, and cleaned each cube to $3\times$ this level. The pixel size for each cube was computed to be $\sim4\times$ smaller than the synthesized beam size in order to properly sample the beam. 

All target sources were unresolved by our observations, except in two cases: J17104, and J17251. When unresolved, the final spectrum was extracted from the central pixel (i.e., pixel of maximum flux density) of each cleaned data cube. For the two resolved sources, we extracted two spectra each from the two local maxima in flux density. To compute the absorption profile (\thi), we normalized each extracted spectrum by the flux density at the same pixel in the continuum image.

In Table~\ref{t:mach_obs} we list the names, coordinates, flux densities \citep{becker1995}, and observation details (array configuration, observation date, on-source time) for the MACH targets. The \thi\ spectra are shown in Figure~\ref{f:spectra_summary}, and Figure~\ref{f:position_comparison} displays the target locations overlaid on an \hi\ column density map from the EBHIS survey \citep{winkel2016}.

\subsection{Matching HI Emission}

In addition to absorption, measuring the physical properties of neutral ISM structures requires constraints for \hi\ emission. 

\begin{deluxetable*}{l|cc|c|ccc|l}
\tablecaption{\label{t:mach_obs} MACH Observation Information}
\tablehead{
\colhead{Name} & \colhead{RA}    & \colhead{Dec} &  \colhead{$S_{1.4\rm GHz,\,FIRST}$} & \colhead{VLA Config} & \colhead{Date} & \colhead{Time on source} & \colhead{$\sigma_{\tau_{\rm HI}}$}    
\\ 
\colhead{} & \colhead{($^{\circ}$)} & \colhead{($^{\circ}$)} &  \colhead{(mJy/bm)} & \colhead{} & \colhead{} & \colhead{(min)} & \colhead{} \\ 
\colhead{(1)} & \colhead{(2)} & \colhead{(3)} &  \colhead{(4)} & \colhead{(5)} & \colhead{(6)} & \colhead{(7)} & \colhead{(8)}  
} 
\startdata
J140028+621038  &  210.118  &  62.177  &  4256.2  &  C  &  17-Aug-21  &  20.0  &  0.002  \\ 
J143645+633638  &  219.189  &  63.610  &  855.4  &  C  &  17-Aug-19  &  28.6  &  0.004  \\ 
J143844+621154  &  219.685  &  62.198  &  2318.6  &  C  &  17-Aug-21  &  20.0  &  0.002  \\ 
J144343+503431  &  220.930  &  50.575  &  1181.6  &  BnA  &  18-Feb-16  &  25.4  &  0.002  \\ 
J145107+415441  &  222.780  &  41.911  &  792.2  &  B  &  18-Jan-14  &  30.6  &  0.005  \\ 
J145408+500331  &  223.535  &  50.058  &  868.2  &  BnA  &  18-Feb-01  &  27.2  &  0.006  \\ 
J150409+600055  &  226.037  &  60.015  &  1494.9  &  C  &  17-Aug-23  &  22.2  &  0.003  \\ 
J150757+621334  &  226.987  &  62.226  &  511.4  &  C  &  17-Aug-18  &  72.1  &  0.006  \\ 
J151020+524430  &  227.584  &  52.742  &  500.5  &  BnA  &  18-Feb-01  &  60.2  &  0.007  \\ 
J153948+611356  &  234.949  &  61.232  &  485.1  &  C  &  17-Aug-26  &  74.1  &  0.008  \\ 
J154122+382029  &  235.345  &  38.341  &  564.4  &  B  &  17-Dec-14  &  50.1  &  0.006  \\ 
J155931+434916  &  236.285  &  47.865  &  739.9  &  B  &  18-Jan-07  &  33.4  &  0.007  \\ 
J154525+462244  &  236.356  &  46.379  &  459.0  &  BnA  &  18-Feb-07  &  29.2  &  0.010  \\ 
J154840+614731  &  237.167  &  61.792  &  527.6  &  C, C$\rightarrow$B  &  17-Aug-26  &  74.3  &  0.004  \\ 
J155128+640537  &  237.866  &  64.093  &  662.9  &  C  &  17-Aug-22  &  68.4  &  0.008  \\ 
J160246+524358  &  240.693  &  52.733  &  557.5  &  B, BnA  &  18-Jan-28  &  58.6  &  0.005  \\ 
J160427+605055  &  241.113  &  60.848  &  572.2  &  C  &  17-Aug-26  &  64.1  &  0.008  \\ 
J161148+404020  &  242.952  &  40.672  &  553.5  &  B  &  17-Nov-17  &  51.0  &  0.005  \\ 
J162557+413440  &  246.490  &  41.578  &  1694.6  &  B  &  17-Oct-20  &  22.3  &  0.003  \\ 
J163113+434840  &  247.804  &  43.811  &  581.0  &  B  &  17-Nov-12  &  24.5  &  0.009  \\ 
J163433+624535  &  248.638  &  62.759  &  4829.3  &  C  &  17-Aug-24  &  13.0  &  0.001  \\ 
J163510+584837  &  248.793  &  58.810  &  506.2  &  C$\rightarrow$B  &  17-Aug-29  &  64.1  &  0.005  \\ 
J164032+382641  &  250.133  &  38.445  &  464.3  &  B  &  18-Jan-27  &  52.5  &  0.006  \\ 
J164258+394837  &  250.745  &  39.810  &  6050.1  &  B  &  18-Jan-05  &  20.2  &  0.001  \\ 
J164800+374429  &  252.000  &  37.741  &  625.7  &  BnA$\rightarrow$A  &  18-Feb-21  &  26.1  &  0.002  \\ 
J165352+394536  &  253.467  &  39.760  &  1394.4  &  B  &  17-Dec-05  &  24.1  &  0.004  \\ 
J165720+570553  &  254.336  &  57.098  &  813.5  &  BnA  &  18-Feb-02  &  30.1  &  0.002  \\ 
J165746+480832  &  254.445  &  48.142  &  981.9  &  B  &  18-Jan-17  &  34.3  &  0.002  \\ 
J165802+473749  &  254.511  &  47.630  &  873.8  &  BnA  &  18-Feb-02  &  29.3  &  0.005  \\ 
J165822+390625  &  254.592  &  39.107  &  646.7  &  B  &  18-Jan-11  &  43.2  &  0.006  \\ 
J170246+551639  &  255.696  &  55.277  &  574.3  &  B  &  18-Jan-26  &  59.1  &  0.003  \\ 
J170253+501741  &  255.721  &  50.295  &  951.5  &  B  &  18-Jan-06  &  64.8  &  0.006  \\ 
J170541+521454  &  256.422  &  52.249  &  504.0  &  BnA$\rightarrow$A  &  18-Feb-26  &  29.3  &  0.007  \\ 
J171044+460124  &  257.683  &  46.026  &  643.5  &  B  &  17-Dec-11  &  33.3  &  0.008  \\ 
J171044+460124  &  257.687  &  46.025  &  985.0  &   ''  &  ''   &  ''   &  0.006  \\ 
J171959+640436  &  259.997  &  64.076  &  817.9  &  C  &  17-Aug-26  &  28.6  &  0.007  \\ 
J172339+523648  &  260.916  &  52.613  &  465.4  &  B  &  18-Jan-28  &  58.5  &  0.010  \\ 
J172516+403641  &  261.317  &  40.611  &  635.6  &  B, BnA, BnA$\rightarrow$A  &  18-Jan-28  &  87.1  &  0.004  \\ 
J172516+403641  &  261.319  &  40.612  &  635.6  &  ''   &   ''  &  ''   &  0.008  \\ 
J173044+490626  &  262.685  &  49.107  &  782.2  &  B  &  17-Dec-21  &  37.1  &  0.005  \\ 
J173054+381150  &  262.725  &  38.197  &  530.0  &  B  &  17-Dec-03  &  52.6  &  0.008  \\ 
J173957+473758  &  264.988  &  47.633  &  907.9  &  B  &  18-Jan-19  &  34.3  &  0.005  \\ 
J174036+521143  &  265.154  &  52.195  &  1508.2  &  B  &  18-Jan-28  &  24.2  &  0.003  \\ 
J174223+540332  &  265.598  &  54.059  &  450.2  &  B  &  17-Nov-25  &  58.6  &  0.009  \\ 
\enddata
\tablecomments{MACH sources listed in order of RA. (1): Target name; (2, 3): RA and Dec coordinates; (4): Flux density at $21\rm\,cm$ from the VLA-FIRST survey \citep{becker1995}; (5): VLA array configuration(s) (``$\rightarrow$" denotes configuration move); (6): Observation date; (7) Total time on-source; (8): Achieved. RMS noise in optical depth ($\sigma_{\tau_{\rm HI}}$) per $0.42\rm\,km\,s^{-1}$ channels.}
\end{deluxetable*}

This is not straightforward to acquire, as the presence of the background continuum source precludes us from observing emission from precisely the same structures as we are sensitive to in absorption. Furthermore, obtaining emission observations on the same angular scale as absorption from a facility such as the VLA is prohibitively expensive, as the noise in brightness temperature is proportional to the inverse square of the half power at full width of the telescope beam \citep{dickey1990}. 

A common strategy for estimating the ``expected" \hi\ brightness temperature profile in the absence of background continuum therefore is to observe positions surrounding the target source with lower-resolution single-dish telescopes and interpolate between them. The highest-resolution survey of $21\rm\,cm$ emission to date in the MACH footprint is from the Effelsberg-Bonn \hi\ Survey \citep[EBHIS;][]{winkel2010, kerp2011, winkel2016}. EBHIS is an all-Northern sky (North of Dec$=-5^{\circ}$) survey with high angular resolution ($9^\prime$), high sensitivity ($\sigma_{\rm rms}=90\rm\,mK$) and $1.3\rm\,km\,s^{-1}$ per channel velocity resolution.\footnote{We note that the coarser velocity resolution of EBHIS relative to MACH (i.e., $1.3\rm\,km\,s^{-1}$ vs. $0.42\rm\,km\,s^{-1}$) does not significantly affect our results. The typical CNM line width is $\gtrsim 2\rm\,km\,s^{-1}$ \citep{murray2018b} and all spectra are re-sampled to $0.1\rm\,km\,s^{-1}$ per channel resolution prior to fitting to avoid aliasing narrow components. }

For each MACH target, we extract all EBHIS spectra within a circle of radius $12^{\prime}$ ($4$ pixels, where each pixel corresponds to $3^{\prime}$), and exclude spectra within a circle corresponding roughly to one EBHIS beam full width at half maximum (FWHM) from the target (radius $1.5$ pixels), which are typically contaminated by the continuum source. Instead of interpolating between the resulting set of \noff\ $T_B(v)$ spectra, we will use each of these spectra separately as if it were the on-source spectrum, and use the distributions in the resulting fitted parameters to incorporate the significant uncertainty in $T_B(v)$ variations to infer \hi\ physical properties. 

\subsection{Uncertainty}

The brightness temperature of \hi\ at Galactic velocities can raise the system temperature of a radio receiver, and therefore the uncertainty in \thi\ will vary as a function of velocity. To estimate the uncertainty spectra ($\sigma_{\tau\rm HI}(v)$) we follow the methods outlined in \citet[][Section 3.2]{murray2015}, which were developed following \citep{roy2013}. To estimate the uncertainty in the associated brightness temperature ($\sigma_{TB}(v)$), we compute the standard deviation of $T_B(v)$ from all \noff\ off-positions extracted around each target source. 
Figure~\ref{f:spectra_summary} displays the $44$ MACH sightlines, including \thi, and $T_B(v)$ for the \noff\ off-positions for each source. In Table~\ref{t:mach_obs} we include the median root mean square (rms) uncertainty in optical depth per $0.42\rm\,km\,s^{-1}$ channels for all LOS. 

All MACH \thi\ spectra and their associated uncertainties are publicly available.\footnote{DOI link active upon publication: https://doi.org/10.7910/DVN/QVYLDV}

\section{Analysis}
\label{sec:analysis}

\subsection{Line of Sight Properties}

To estimate the ensemble properties of \hi\ for the MACH and comparison samples, we integrate \thi\ and $T_B(v)$ along the sightline. 

The total column density (\nhi) is given by,

\begin{equation}
    N({\rm HI}) = C_0 \int \tau_{\rm HI}\,T_s\,dv,
\end{equation}

\noindent where $C_0=1.823\times10^{18}\rm\,cm^{-2}/(K\,km\,s^{-1})$ \citep{draine2011} and $T_s$ is the \hi\ excitation temperature, also known as the ``spin" temperature. To approximate the spin temperature using observable \hi\ properties, we assume that \hi\ a at a single temperature dominates each velocity channel, so that,

\begin{equation}
    T_s(v) \simeq \frac{T_B(v)}{1-e^{-\tau_{\rm HI}(v)}}.
\end{equation}

\noindent As a result, the \nhi\ is given by \citep[e.g.,][]{dickey1982},

\begin{equation}
    N({\rm HI}) \simeq C_0 \int \frac{\tau_{\rm HI}\,T_B}{(1-e^{-\tau_{\rm HI}})}\,dv.
\label{e:nhi}
\end{equation}

\noindent The approximation to \nhi\ given by Equation~\ref{e:nhi} has been shown to agree with sophisticated multiphase analysis of $21\rm\,cm$ spectral line pairs \citep{stanimirovic2014, murray2018b}. Specifically, in low-column density regimes (\nhi$<5\times10^{20}\rm\,cm^{-2}$), Equation~\ref{e:nhi} is fully consistent with the results of decomposing $T_B(v)$ and \thi\ into individual velocity components of distinct temperature and density and accounting for the order of components along the sightline. In addition, \citet{kim2014} showed using synthetic $21\rm\,cm$ observations of 3D hydrodynamic simulations of the Galactic ISM that Equation~\ref{e:nhi} approximates the true simulated column density to within $5\%$. For our high-latitude samples (MACH and comparison, all with $|b|>10^{\circ}$), we will use this approximation for \nhi.

If the gas is optically-thin ($\tau_{\rm HI}<<1$), Equation~\ref{e:nhi} reduces to,

\begin{equation}
    N({\rm HI})^* = C_0 \int T_B \,dv,
\end{equation}

\begin{figure*}
\begin{center}
\vspace{-60pt}
\includegraphics[width=\textwidth]{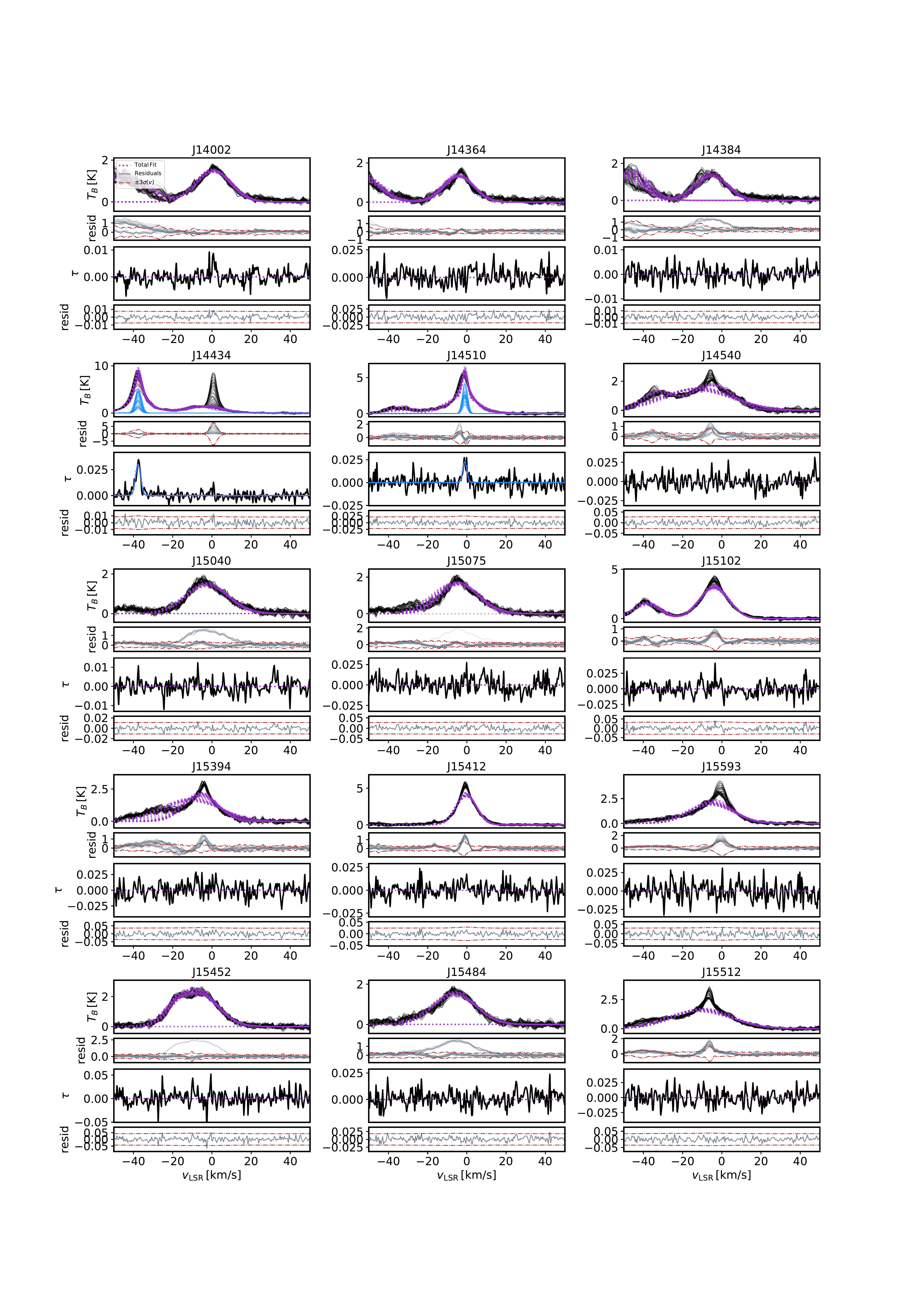}
% \vspace{-70pt}
\caption{Gaussian fits to MACH LOS using GaussPy. Each panel corresponds to a different sightline, including $T_B(v)$ from the \noff\ off-positions (top sub-panel), and \thi\ (bottom sub-panel), with the total fits (dotted purple) to \thi\ \citep[Equation 1;][]{murray2018b} and $T_B(v)$ (Equation~\ref{e:tbfit}), the individual components (solid, colors) and the residuals from the fits (sub-panels; grey), along with $\pm3\sigma$ uncertainties (red).}
\label{f:mach_decomposition}
\end{center}
\end{figure*}

\begin{figure*}
% \ContinuedFloat
\begin{center}
\vspace{-60pt}
\includegraphics[width=\textwidth]{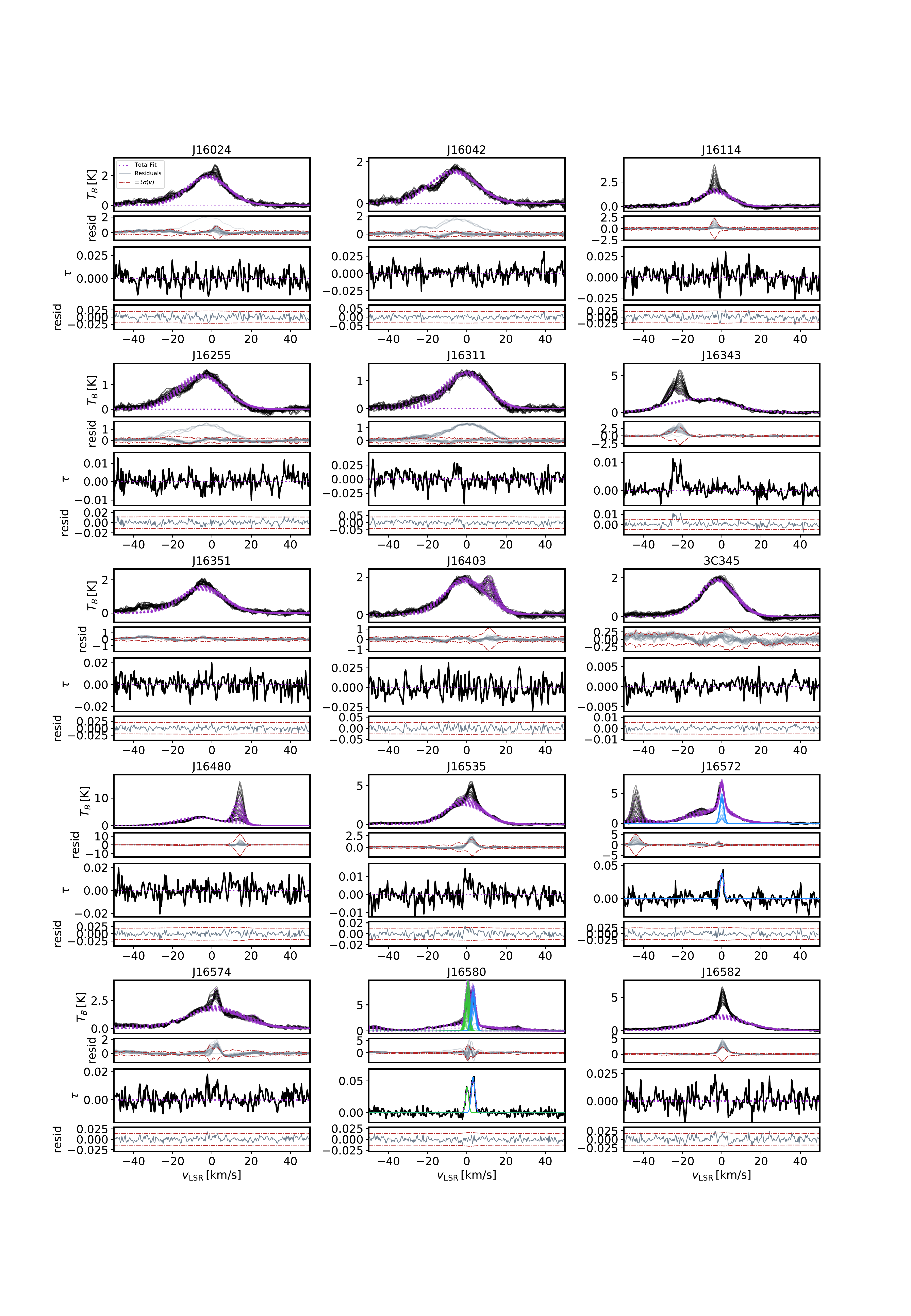}
% \vspace{-70pt}
% \caption{(contd.) Gaussian fits to MACH LOS using GaussPy. Each panel corresponds to a different sightline, including $T_B(v)$ from the \noff\ off-positions (top sub-panel), and \thi\ (bottom sub-panel), with the total fits (dotted purple) to \thi\ \citep[Equation 1;][]{murray2018b} and $T_B(v)$ (Equation~\ref{e:tbfit}), the individual components (solid, colors) and the residuals from the fits (sub-panels; grey), along with $\pm3\sigma$ uncertainties (red).}
% \label{f:mach_decomposition_b}
\end{center}
\end{figure*}

\begin{figure*}
% \ContinuedFloat
\begin{center}
\vspace{-60pt}
\includegraphics[width=\textwidth]{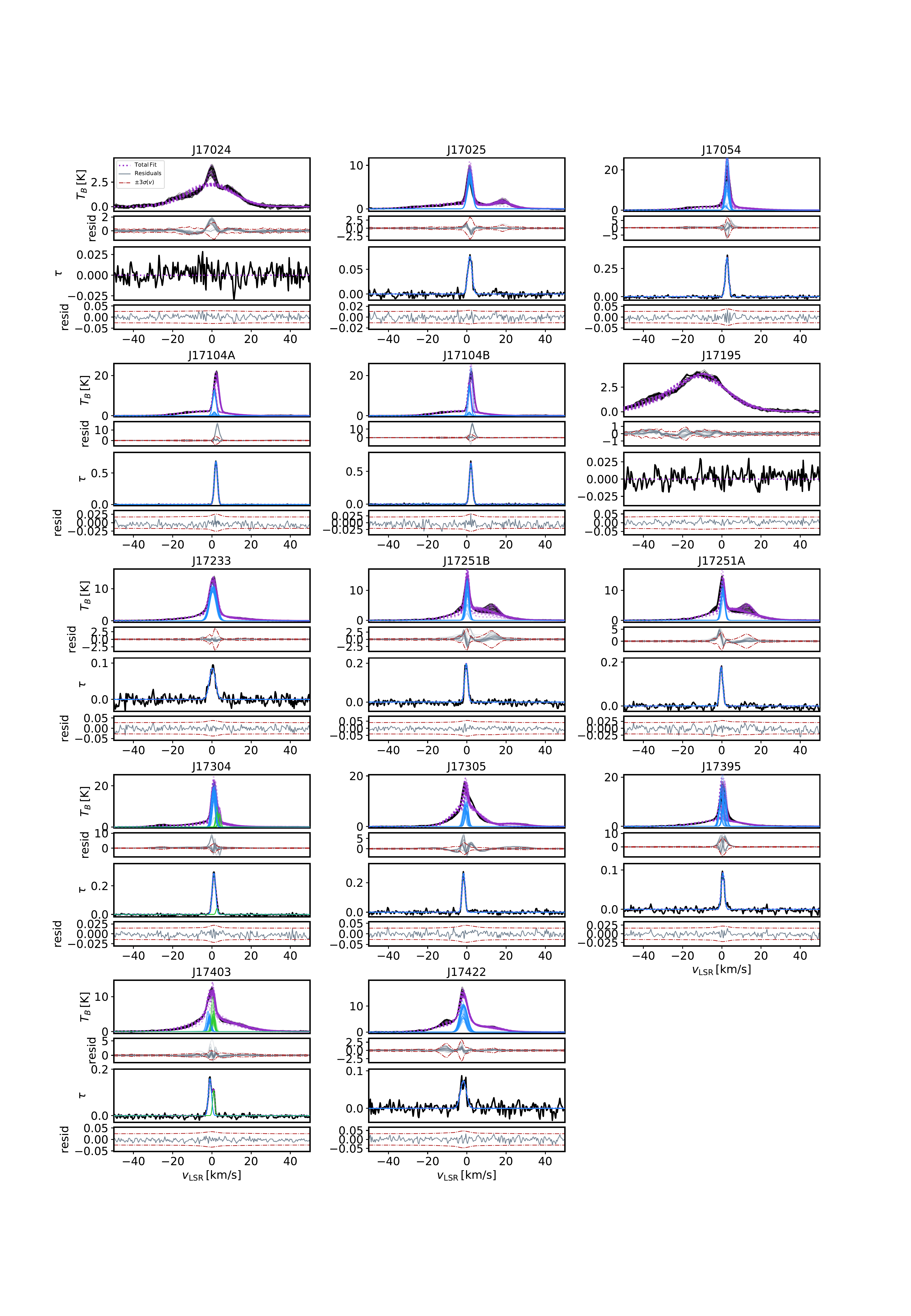}
% \vspace{-70pt}
% \caption{(contd.) Gaussian fits to MACH LOS using GaussPy. Each panel corresponds to a different sightline, including $T_B(v)$ from the \noff\ off-positions (top panel), and \thi\ (bottom panel), with the total fits (dotted purple) to \thi\ \citep[Equation 1;][]{murray2018b} and $T_B(v)$ (Equation~\ref{e:tbfit}), the individual components (solid, colors) and the residuals from the fits (sub-panels; grey), along with $\pm3\sigma$ uncertainties (red).}
% \label{f:mach_decomposition}
\end{center}
\end{figure*}

\noindent which is a common assumption used to compute \nhi\ in the absence of \thi\ measurements. To quantify how much column density is ``missed" in the optically-thin limit due to the presence of optically-thick \hi, we compute the ratio of the two column density estimates,

\begin{equation}
    \mathcal{R}_{\rm HI} = \frac{N({\rm HI})}{N({\rm HI})^*}.
    \label{e:rhi}
\end{equation}

Next, we estimate the relative contribution of cold vs. warm \hi\ by computing the fraction of the CNM along the line of sight (\fcnm). We follow the methods outlined by \citet{murray2020} \citep[their Section 2.5.2; based on][]{kim2014}, who argue that \fcnm\ is approximated by,

\begin{equation}
    f_{\rm CNM} \approx \frac{T_c}{\langle T_s \rangle} \frac{T_{s,w}-\langle T_s \rangle}{T_{s,w} - T_c},
    \label{e:fcnm}
\end{equation}

\noindent where $\langle T_s \rangle$ is the optical depth-weighted average spin temperature,

\begin{equation}
    \langle T_s \rangle = \frac{\int \tau_{\rm HI}\,T_s\,dv}{\int \tau_{\rm HI}\,dv},
\end{equation}

\noindent $T_c$ is the kinetic temperature of the CNM and $T_{s,w}$ is the spin temperature of the WNM. Following \citet{murray2020} we set $T_c=50\rm\,K$ and $T_{s,w}=1500\rm\,K$. To account for the considerable uncertainty in these estimates (e.g., the true WNM $T_s$ can be much higher than $1500\rm\,K$), we vary these estimates between $20<T_c<150\rm\,K$ \citep{dickey2000} and $1000<T_{s,w}<6000\rm\,K$ when computing the uncertainties for \fcnm\ (see below). 
To compute the uncertainties in \nhi, \rhi\ and \fcnm, we perform a simple Monte Carlo exercise. Over $10^5$ trials, we recompute each value after adding random noise to $T_B(v)$ and \thi\ drawn from $\pm 3\times \sigma_{TB}(v)$ and $\pm 3 \times \sigma_{\tau\rm HI}(v)$ respectively. For \fcnm, we also vary the values of $T_c$ and $T_{s,w}$ as discussed above. We then repeat this computation for each of the \noff\ off-positions. The final values and uncertainties for each parameter are computed as the median and standard deviation over all trials. 

\subsection{Gaussian Decomposition}
\label{sec:gauss_fits}

Beyond integrated properties, we are interested in estimating the properties of individual \hi\ structures. To decompose the $21\rm\,cm$ spectral line pairs, we follow the methods described by \citet{murray2018b} for the 21-SPONGE survey (summarized here for clarity), which are based on the strategy employed by \citet[][; hereafter HT03]{heiles2003, heiles2003b}. 

We begin by decomposing each \thi\ spectrum using the Autonomous Gaussian Decomposition (AGD) algorithm \citep{lindner2015}, implemented via its open-source Python package GaussPy\footnote{\url{https://github.com/gausspy/gausspy}}. AGD provides initial guesses for the number and properties of all Gaussian components (amplitude ($\tau_0$), mean velocity ($v_0$) and linewidth (FWHM $\delta v$)) within \thi\ by computing successive numerical derivatives with regularization. First, all $21\rm\,cm$ emission/absorption pairs are re-sampled to $0.1\rm\,km\,s^{-1}$ per channel resolution to avoid aliasing narrow components \citep{lindner2015}. For the fit, we use the ``two-phase" implementation of AGD, wherein we identify CNM-like (narrow linewidth) and WNM-like (broad linewidth) components in two steps, using regularization parameters $\alpha_1=1.12$ and $\alpha_2=2.75$ and a signal-to-noise cutoff of $S/N>3$ for both phases \citep{murray2017, murray2018b}. Given $J$ Gaussian components predicted by AGD for each spectrum, we produce a model for \thi\ using \citet{murray2018b} Equation 1 via least-squares fit implemented in GaussPy. 

The next step is to determine the emitting properties of the fitted absorption components. Specifically, we assume that the $J$ fitted absorption components contribute both optical depth and emission along the line of sight and an additional $K$ components, dominated by WNM, are only detected in emission \citep[e.g., ][HT03]{mebold1997, dickey2000, murray2015}. First,  we use the Levenberg-Marquart algorithm implemented in the Python package lmfit\footnote{For this work, we used lmfit version 1.0.1.} to perform a least-squares fit of the fitted properties of the $J$ absorption components to $T_B(v)$. We allow the amplitudes of the components to vary freely, and constrain their mean velocities to vary with $\pm2$ channels, and their FWHM to vary by $\pm10\%$. From the residuals of this initial fit, we use AGD to determine starting guesses for the properties of $K$ additional emission-only components using the one-phase fit \citep[$\alpha=3.75,\,S/N\ge3$;][]{murray2018b}. The properties of all $J+K$ components are estimated with an additional least-squares fit to $T_B(v)$ \citep[][Equation 2]{murray2018b}. 

We emphasize that the GaussPy fits are sensitive to the selection of the regularization and other parameters (e.g., signal to noise thresholds, mean velocity and FWHM variation). The systematic uncertainty in the resulting parameters is therefore large, and similar to if we selected the Gaussian fit properties by hand as is traditionally done. The benefit of the GaussPy implementation is that the results are reproducible. 

In summary, for each source we have constraints for the properties for the $J$ components fitted to \thi: including their amplitudes ($\tau_{0,j}$), full widths at half maximum ($\delta v_{0,j}$) and mean velocities ($v_{0,j}$) in absorption.  For each of the \noff\ off-positions in $T_B(v)$, we have constraints for the amplitudes ($T_{0,j}$), FWHMs ($\delta v_{j,\rm em}$) and mean velocities ($v_{0,j,{\rm em}}$) of the $J$ absorption-detected components in emission, as well as the  amplitudes ($T_{0,k}$), FWHMs ($\delta v_{k}$) and mean velocities ($v_{0,k}$) of the $K$ additional components fitted only in emission. 
\begin{figure*}
\begin{center}
% \vspace{-80pt}
\includegraphics[width=\textwidth]{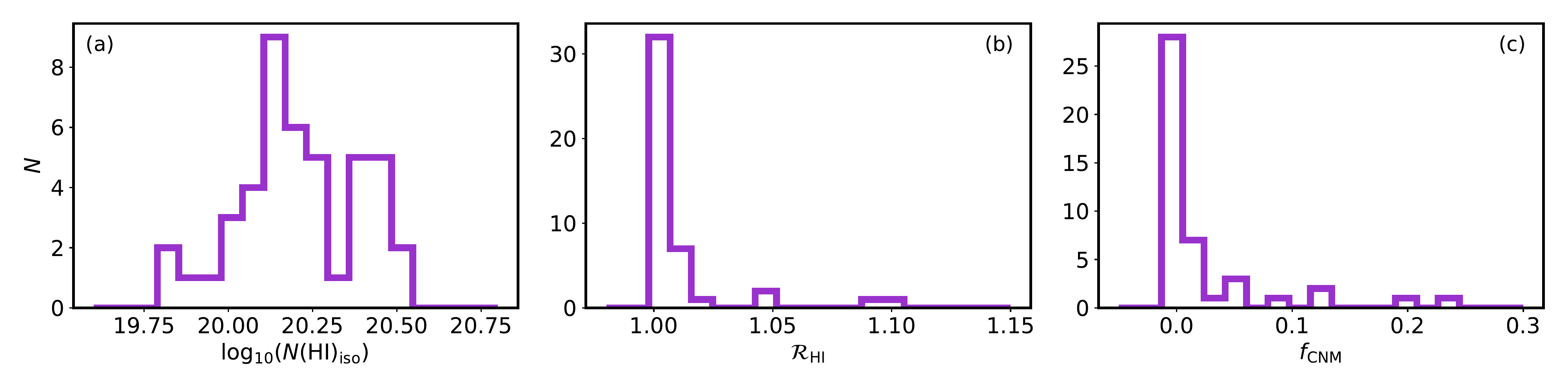}
% \vspace{-80pt}
\caption{Histograms of integrated properties along MACH LOS, including total \hi\ column density ($N({\rm HI})$; Equation~\ref{e:nhi}; a), optical depth correction factor ($\mathcal{R}_{\rm HI}$; Equation~\ref{e:rhi}; b) and CNM fraction ($f_{\rm CNM}$; Equation~\ref{e:fcnm}; c).}
\label{f:integrated_properties}
\end{center}
\end{figure*}

\begin{figure*}
\begin{center}
\includegraphics[width=\textwidth]{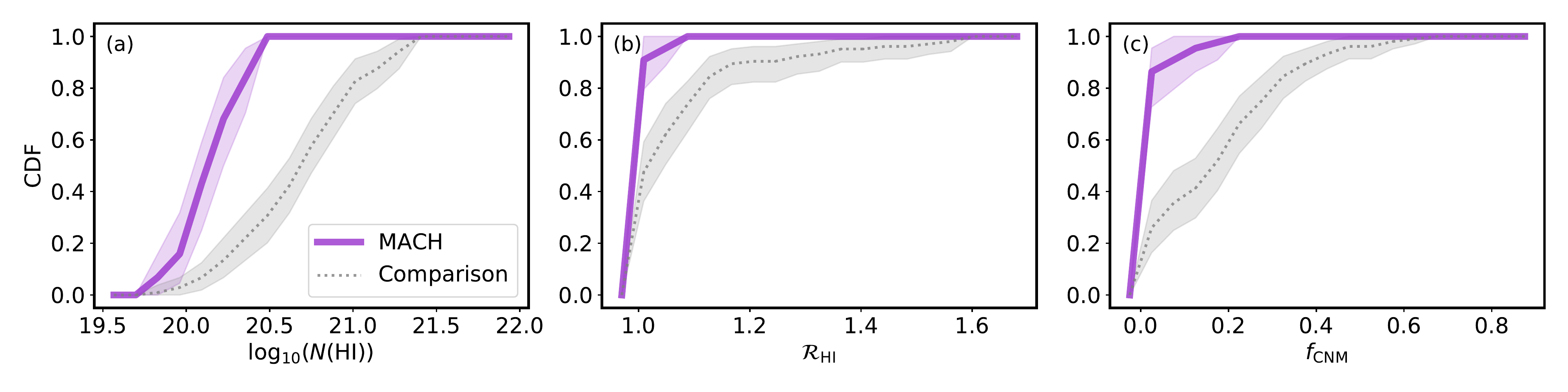}
\caption{Comparison of cumulative distribution functions (CDFs) of integrated \hi\ properties between the MACH sample (purple) and the high-latitude comparison sample (grey; Section~\ref{sec:comparison}), including total \hi\ column density ($N({\rm HI})$; Equation~\ref{e:nhi}; a),  optical depth correction factor ($\mathcal{R}_{\rm HI}$; Equation~\ref{e:rhi}; b) and CNM fraction ($f_{\rm CNM}$; Equation~\ref{e:fcnm}; c). The uncertainty ranges in the CDFs are computed by bootstrapping the samples (see text).}
\label{f:integrated_param_comparison}
\end{center}
\end{figure*}

\subsubsection{Inferring physical properties}

Our ultimate goal is to infer important physical properties such as kinetic temperature ($T_k$), spin temperature ($T_s$), column density ($N(\rm HI)$) and turbulent mach number ($\mathcal{M}_t$) using fitted spectral line properties. Following HT03, to this end we need to take into account the order of components along the line of sight ($\mathcal{O}$), as well as the fraction of emission-only components absorbed by foreground absorption components, both of which affect the inferred values of $T_s$ (HT03). In detail we solve,

\begin{equation}
    T_B(v) = T_{B,\rm abs}(v) + T_{B,\rm em}(v),
    \label{e:tbfit}
\end{equation}

\noindent where $T_{B,\rm abs}(v)$ and $T_{B,\rm em}(v)$ are the contribution of the $J$ absorption-detected components and $K$ emission-only components to $T_B(v)$ respectively. These are given by,

\begin{equation}
    T_{B,\rm abs}(v) = \sum_{j=0}^{J-1} T_{s,j} (1-e^{-\tau_j(v)})e^{-\sum_{m=0}^{M-1}\tau_m(v)},
     \label{e:tbabs_full}
\end{equation}

\noindent where for each $j^{\rm th}$ component, $\tau_j(v)$ is the Gaussian model in absorption and $T_{s,j}$ is the spin temperature, and the subscript $m$ denotes all components lying \emph{in front} of the $j^{\rm th}$ component along the sightline. The contribution from the $K$ emission-only components is given by, 

\begin{equation}
    T_{B,\rm em}(v) = \sum_{k=0}^{K-1} [\mathcal{F}_k + (1-\mathcal{F}_k)e^{-\tau(v)}] \, g(T_{k}, v_{0,k}, \delta v_k),
    \label{e:tbem_full}
\end{equation}

\noindent where $\mathcal{F}_k$ is the fraction of the $k^{\rm th}$ component which lies in front of the $N$ absorption components and $g(A,v_0,\delta v)$ is a Gaussian function for amplitude ($A$), mean velocity ($v_0$)  and FWHM ($\delta v$). 

Determining the best-fit values of $\mathcal{O}$ and $\mathcal{F}$ for each component requires iterating over all permutations of possible values. Following HT03, we first make several simplifying assumptions. First, for each of the $K$ emission components, we allow $\mathcal{F}_k=(0,0.5,1)$, as finer variations are difficult to distinguish statistically \citep[HT03;][]{murray2015,nguyen2018}. We are then left with $J!\times 3^K$ possible combinations of $\mathcal{O}$ and $\mathcal{F}$ for each sightline. However, in practice, the $\mathcal{O}$ only matters for those which overlap significantly, and therefore we only permute the $J_x\leq J$ components which overlap with at least one other by at least $3\times \sigma_{\tau} \times \delta v_{j,\rm em}$. In addition, we consider $\mathcal{F}$ only for the $K_x\leq K$ emission components which overlap absorption components by at least $3\times \sigma_{T_B}\times \delta {v_k}$. The result is $J_x!\times 3^{K_x}$ iterations per sightline. For MACH, the total iterations ranged from 1 to 54, and for the comparison sample (which includes more complex LOS) the total iterations varied from 1 to 19440.

\begin{figure*}
\begin{center}
\includegraphics[width=\textwidth]{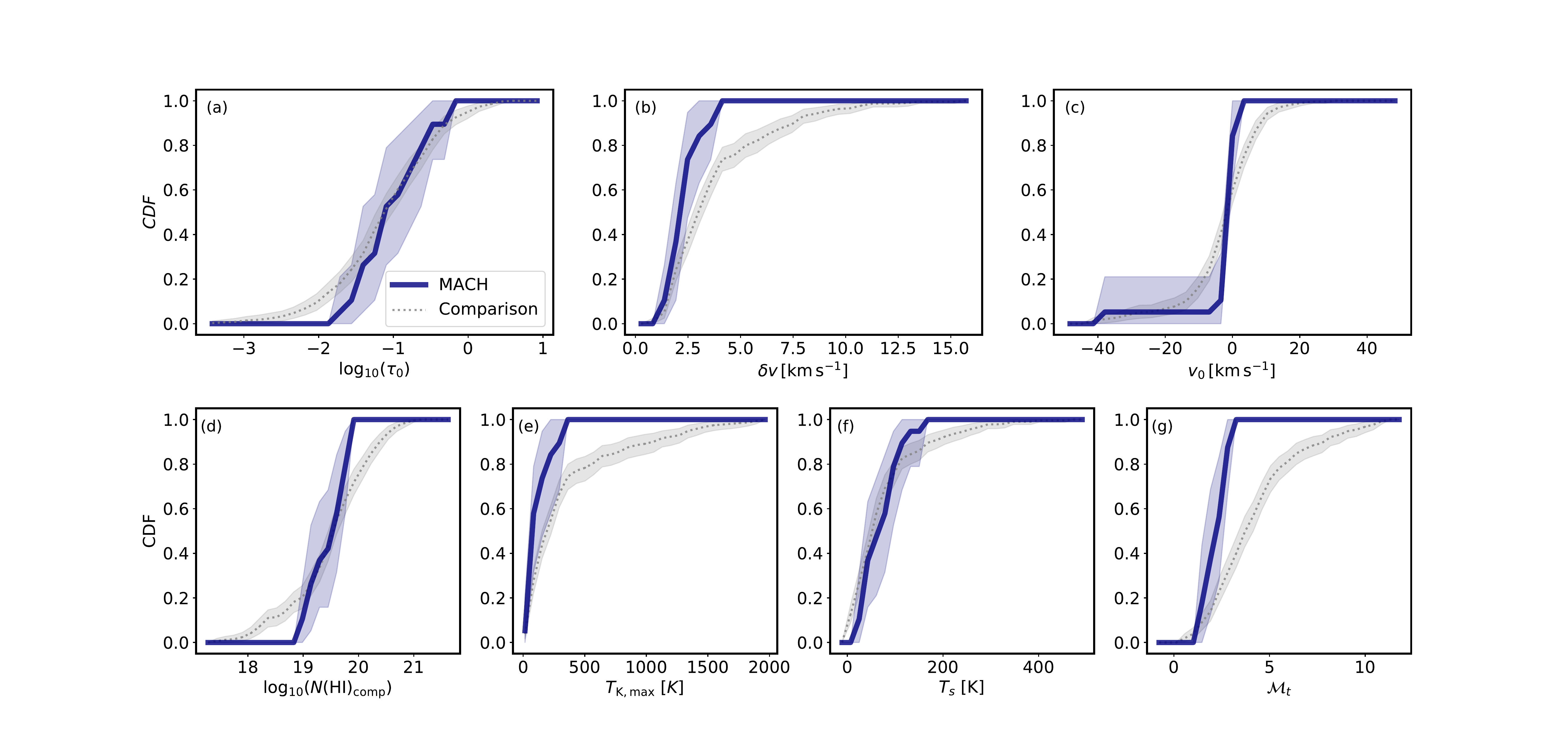}
\caption{Comparison of cumulative distribution functions (CDFs) of parameters from the Gaussian fits to MACH (dark blue) and the high-latitude comparison sample (grey; Section~\ref{sec:comparison}), including amplitude ($\tau_{0}$; a), line width (FWHM; $\delta v$; b), mean velocity ($v_{0}$; c), and derived physical properties including column density (Equation~\ref{e:nhi_comp}, $N({\rm HI})_{\rm comp}$; d), maximum kinetic temperature (Equation~\ref{e:tkmax_comp}, $T_{\rm K, max}$; e), spin temperature ($T_s$; f), and turbulent Mach number (Equation~\ref{e:mt_comp}, $\mathcal{M}_t$; g). The uncertainty ranges in the CDFs are computed by bootstrapping (see text).}
\label{f:fitted_param_comparison}
\end{center}
\end{figure*}

\subsubsection{Final fits}

We repeat this fitting procedure for each of the \noff\ $T_B(v)$ spectra from the off-positions surrounding each absorption target. In each case, we repeat the permutations of $\mathcal{O}$ and $\mathcal{F}$ described above. For each permutation, we estimate $T_{s,j}$ for the $J$ absorption components by least-squares minimization of the fit to Equation~\ref{e:tbfit}. 

Next, we follow HT03 and compute a weighted average of $T_s$ over all permutations of $\mathcal{O}$ and $\mathcal{F}$, where the weight of each trial is the reciprocal of the variance from the residuals to the fit to $T_B(v)$. 
%The best-fit values of $\mathcal{O}$ and $\mathcal{F}$ are selected from the permutation which minimizes the fit residuals across all trials. 

In Figure~\ref{f:mach_decomposition}, we display the results of the fits for all \noff\ off-positions for all MACH LOS. We observe that the fit to \thi\ performed well, and all residuals are within $\pm3\sigma_{\tau \rm HI}(v)$.
%In addition, all detected absorption components are successfully recovered in the fit to $T_B(v)$. 
We also observe that the overall emission fits exhibit strong residuals. These generally fall within the $\pm3\sigma_{T_B}(v)$ uncertainties, which are considerable due to the strong variation in $T_B(v)$ between off positions. In addition, by design, after accounting for detected absorption components the fit to $T_B(v)$ is sensitive only to broad, WNM-like components parameterized by a single ``one-phase" regularization parameter via AGD. So, not only is the procedure trained against fitting additional, CNM-like narrow components to $T_B(v)$, but narrow emission features also correspond to the strongest per-channel uncertainties in $T_B(v)$ (Figure~\ref{f:spectra_summary}), making them even less likely to be included. As we are chiefly concerned with the properties of the absorption-detected components, and considering the well-known uncertainties of matching pencil-beam absorption measurements with emission derived from significantly larger angular scales, we accept the increased uncertainties in the emission fits presented here, which will propagate into the uncertainties in the derived parameters. 

Overall, following the fits, for each of the $J$ detected absorption components we have \noff\ estimates of $T_{s,j}$ and the Gaussian parameters in emission ($T_{0,j}$, $\delta v_{j,\rm em}$, $v_{0,j,{\rm em}}$). The final fitted values for these properties are computed by bootstrapping the \noff\ values with replacement over $10^4$ trials, and computing the mean of the resulting distribution. The final uncertainties are computed by adding the standard deviation of the bootstrapped distribution (i.e., the uncertainty due to the variation off positions) in quadrature with the mean uncertainty from the least-squares fit to $T_B(v)$ over all \noff\ positions. 

Given $T_s$ for each absorption component, we compute the column density per component as,

\begin{equation}
    N({\rm HI})_{\rm abs} = C_0 \int \tau \,T_s\,dv = 1.064\,C_0\,\tau_0 \,\delta v \,T_s,
        \label{e:nhi_comp}
\end{equation}

\noindent where the factor of $1.064$ converts the product to the area of a Gaussian with the given FWHM and amplitude. We also estimate the maximum kinetic temperature, or the upper limit to the kinetic temperature in the absence of non-thermal broadening, from the absorption line width, via,

\begin{equation}
    T_{k,\rm max} = \frac{m_H}{8\,k_B \ln{2}} \delta v^2 = 21.866 \, \delta v^2
    \label{e:tkmax_comp}
\end{equation}

\noindent for hydrogen mass $m_H$ and Boltzmann's constant $k_B$ \citep{draine2011}. Finally, we compute the turbulent Mach number ($\mathcal{M}_t$) via the ratio of $T_{k,\rm max}$ and $T_s$. Following HT03 (their Equation 17), we compute,

\begin{equation}
    \mathcal{M}_t^2 = 4.2 \left (\frac{T_{k,\rm max}}{T_s}-1 \right ).
    \label{e:mt_comp}
\end{equation}

\subsection{Comparison Sample}
\label{sec:comparison}

To compare \hi\ properties along MACH sightlines with other high-latitude environments, we build a sample of measurements from the literature. We select sources from surveys of \thi\ outside of the Galactic Plane ($|b|>10^{\circ}$).

\begin{enumerate}
    \item \emph{VLA (21-SPONGE)}: The $21\rm\,cm$ Spectral Line Observations of Neutral Gas with the Karl G. Jansky Very Large Array \citep[21-SPONGE;][]{murray2015, murray2018b} is the highest-sensitivity survey for Galactic \thi\ to date at the VLA. The median root mean square (rms) uncertainty in \hi\ optical depth is $\sigma_{\tau HI}\lesssim 0.001$ per $0.42\rm\,km\,s^{-1}$ channels. We select the 44 21-SPONGE spectra with $|b|>10^{\circ}$.
    
    \item \emph{Arecibo (HT03, Perseus)}: We include additional spectra from single-dish surveys of \thi\ at the Arecibo Observatory, including the Millennium Arecibo $21\rm\,cm$ Absorption-Line Survey \citep[HT03;][]{heiles2003b} and a targeted survey of the Perseus molecular cloud and its environment \citep[][hereafter Perseus]{stanimirovic2014, lee2015}. The  median optical depth sensitivity of both surveys $\sigma_{\tau \rm HI}\lesssim0.01$ per $0.18\rm\,km\,s^{-1}$ channels. Although single-dish observations of \thi\ are susceptible to contamination from $21\rm\,cm$ emission within the beam, we find excellent correspondence between these studies and interferometric observations from 21-SPONGE \citep{murray2015}. We select the 60 \thi\ spectra (22 Perseus, 38 HT03) which are unique relative to 21-SPONGE with $|b|>10^{\circ}$.
\end{enumerate}

Figure~\ref{f:position_comparison} includes the locations of the selected targets.

With the comparison sample of \thi\ spectra from 21-SPONGE, Perseus and HT03, we extract \hi\ emission spectra from \noff\ off-positions surrounding each target from EBHIS, and compute corresponding uncertainty spectra ($\sigma_{\tau\rm HI}(v)$ and $\sigma_{TB,\rm exp}(v)$) and integrated properties following the same procedures described above for MACH. 

\begin{figure*}
\begin{center}
\includegraphics[width=\textwidth]{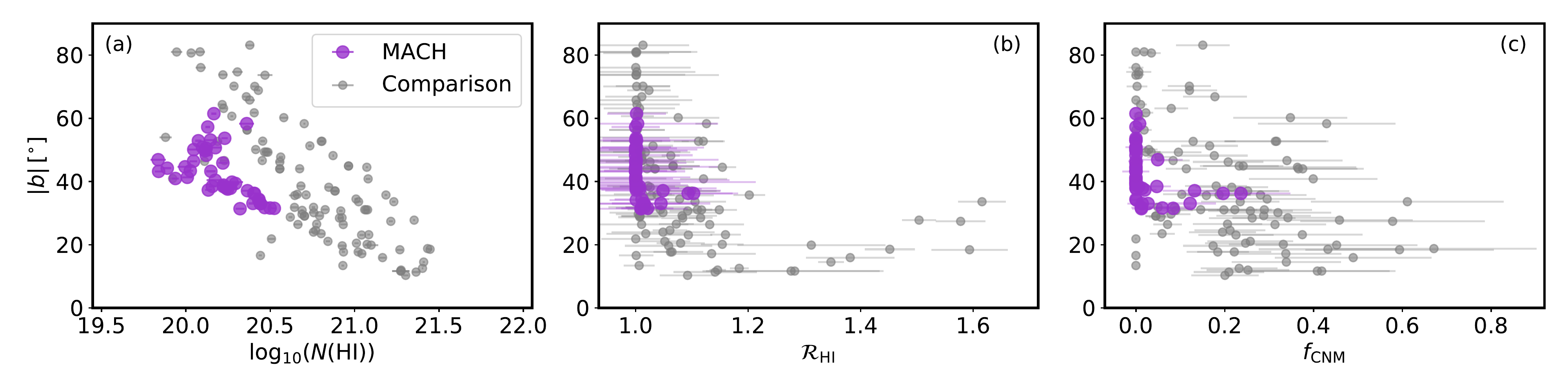}
\caption{Variation in integrated \hi\ properties with absolute Galactic latitude ($|b|$) from the MACH sample (purple) and the high-latitude comparison sample (grey), including total \hi\ column density ($N({\rm HI})$; Equation~\ref{e:nhi}; a),  optical depth correction factor ($\mathcal{R}_{\rm HI}$; Equation~\ref{e:rhi}; b) and CNM fraction ($f_{\rm CNM}$; Equation~\ref{e:fcnm}; c).}
\label{f:integrated_param_latitude}
\end{center}
\end{figure*}

\begin{figure}
\begin{center}
\includegraphics[width=0.45\textwidth]{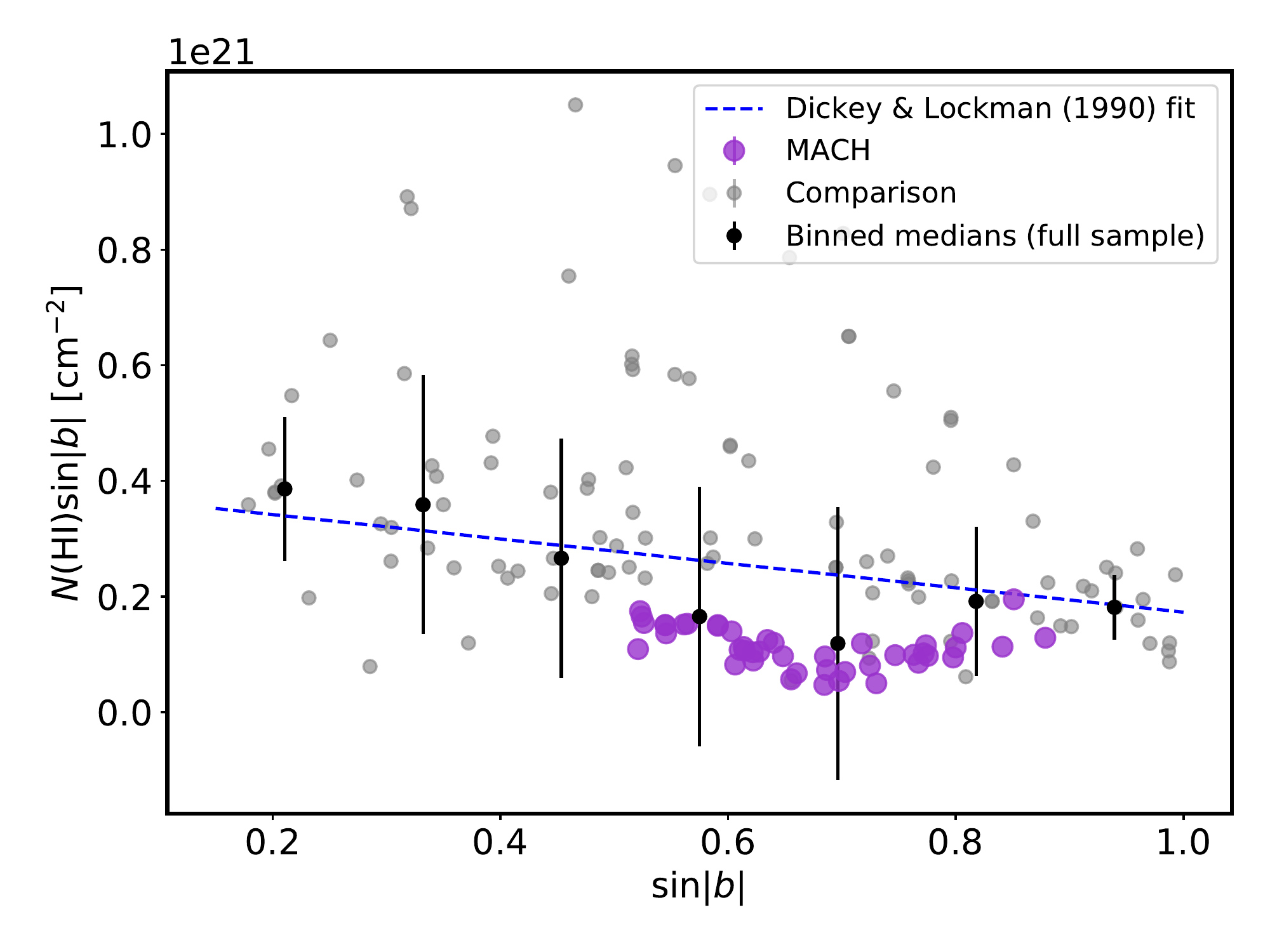}
\caption{Variation in $N({\rm HI})sin|b|$ with $sin|b|$ for this study (MACH (purple) and comparison (grey) LOS). The binned medians and standard deviations of the both samples together (black) are consistent with previous results \citep[e.g.,][]{dickey1990}, and the MACH LOS trace the minimum column densities, emphasizing the departures of the local ISM from plane-parallel symmetry.  }
\label{f:nhisinb}
\end{center}
\end{figure}

\begin{figure*}
\begin{center}
\vspace{-20pt}
\includegraphics[width=\textwidth]{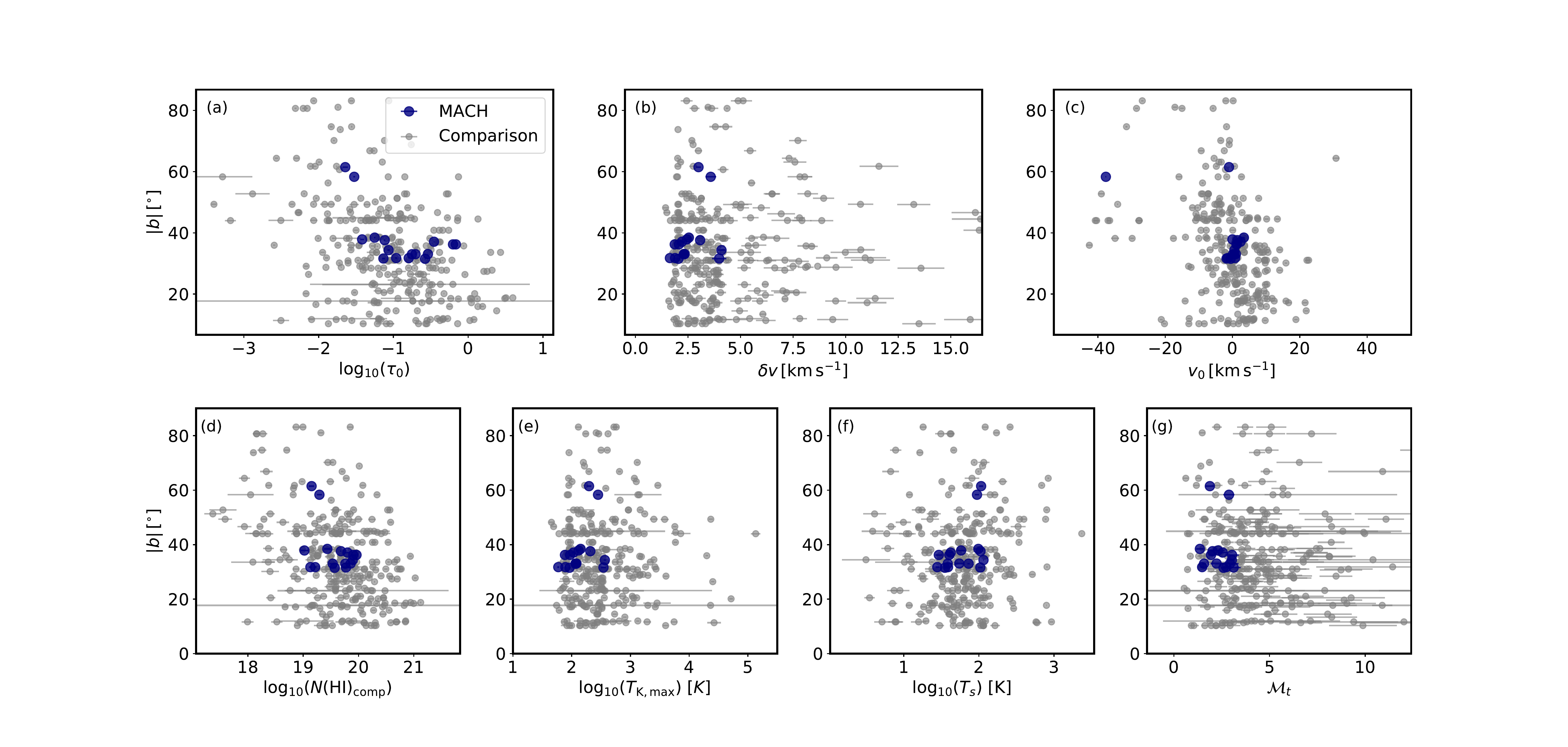}
\vspace{-20pt}
\caption{Variation in component \hi\ properties with absolute Galactic latitude ($|b|$) from MACH (dark blue) and the high-latitude comparison sample (grey), including amplitude ($\tau_{\rm HI}$; a), line width (FWHM; $\delta v$; b), mean velocity ($v_{0}$; c), and derived physical properties including column density (Equation~\ref{e:nhi_comp}, $N({\rm HI})_{\rm comp}$; d), maximum kinetic temperature (Equation~\ref{e:tkmax_comp}, $T_{\rm K, max}$; e), spin temperature ($T_s$; f), and turbulent Mach number (Equation~\ref{e:mt_comp}, $\mathcal{M}_t$; g). }
\label{f:fitted_param_latitude}
\end{center}
\end{figure*}

\begin{figure}
\begin{center}
\includegraphics[width=0.45\textwidth]{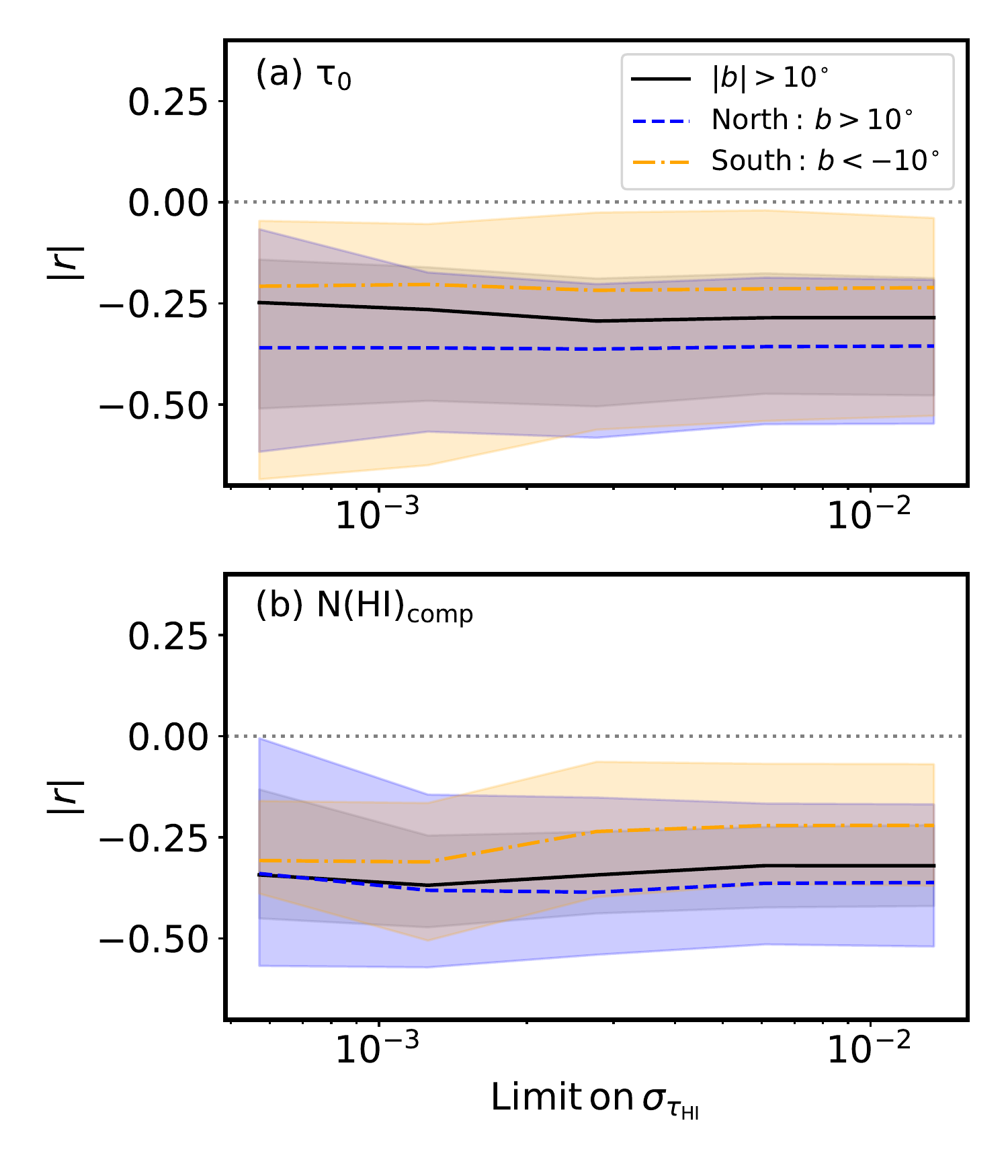}
\vspace{-8pt}
\caption{The effect of optical depth sensitivity on correlation with $|b|$. For $\tau_0$ (a) and  $N({\rm HI})_{\rm comp}$ (b), we plot the $50^{\rm th}$ percentile of the Pearson r coefficient ($|r|$) over $10^4$ block-bootstrapped trials in bins of increasing limits to $\sigma_{\tau_{\rm HI}}$, (only components from LOS with $\sigma_{\tau_{\rm HI}}<3\times$ the limit are included), and repeat for the Northern (blue) and Southern (orange) Galactic hemispheres. }
\label{f:r_sensitivity}
\end{center}
\end{figure}

In addition, we decompose the comparison sample using the same methodology as for MACH. Given that the comparison sample spectra cover an inhomogeneous range of LSR velocity (e.g., SPONGE absorption spectra cover $-50<v<50\rm\,km\,s$ whereas the Arecibo absorption spectra cover $-100<v<100\rm\,km\,s^{-1}$) we first re-sample the comparison sample spectra to the same velocity axis as MACH spectra (i.e., $-150<v<50\rm\,km\,s^{-1}$ with $0.42\rm\,km\,s^{-1}$ channel spacing). For each spectrum, at velocities where there is no absorption coverage by the original observing setup, we add Gaussian noise with amplitude equal to the median RMS noise in off-line channels. Then, as for the MACH sample, we re-sample again to $0.1\rm\,km\,s^{-1}$ per-channel resolution to ensure that we do not alias narrow velocity components \citep{lindner2015}. We emphasize that we will restrict our subsequent analysis of fitted component properties to the velocity range common to all spectra ($-50<v<50\rm\,km\,s^{-1}$).

\begin{figure}
\begin{center}
\includegraphics[width=0.5\textwidth]{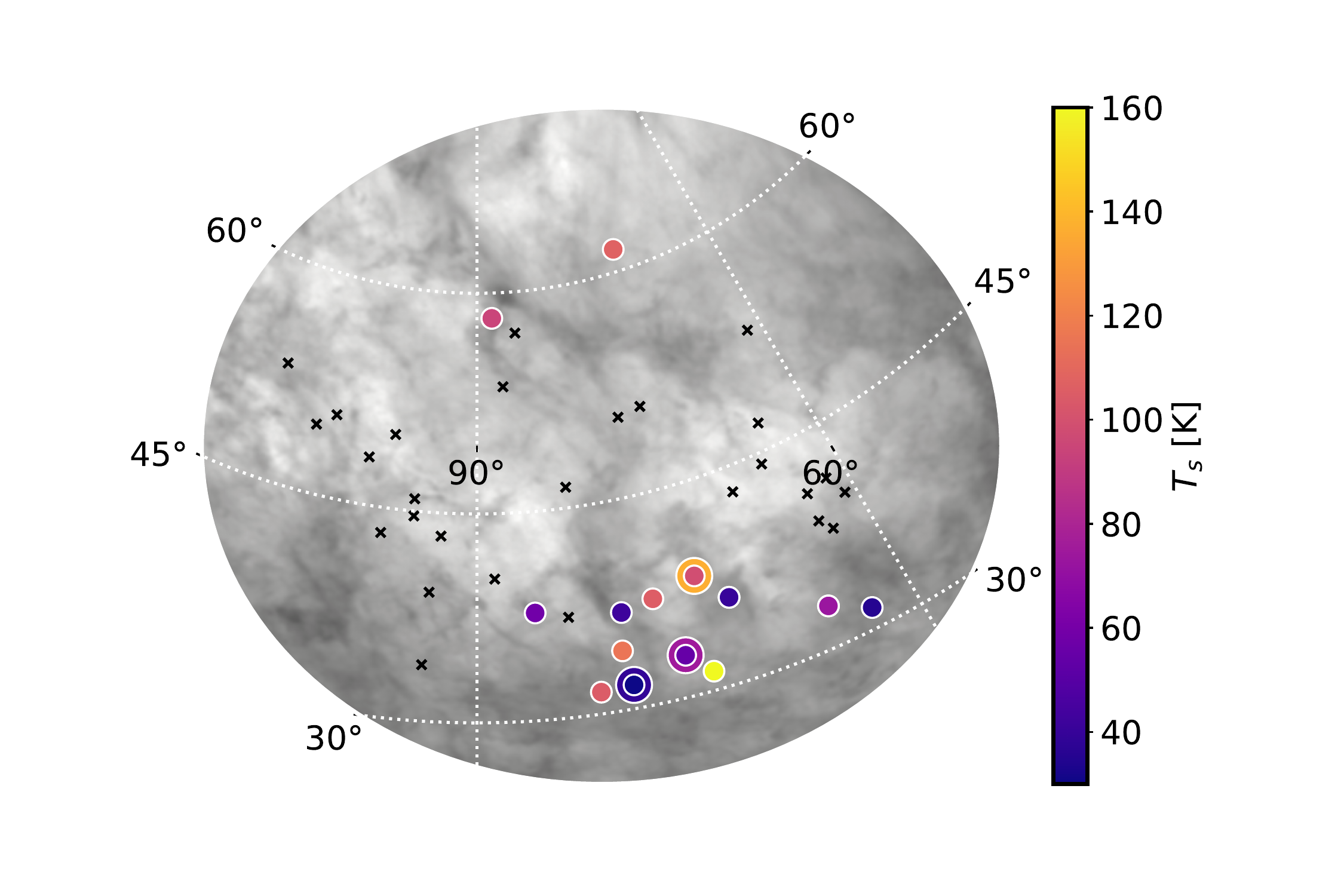}
\caption{Positions of MACH targets overlaid on \nhithin\ \citep[][; same scale as Figure~\ref{f:position_comparison}]{hi4pi2016}. Where an absorption component is detected via GaussPy, the point is colored by inferred spin temperature ($T_s$). For LOS with multiple components, these are shown as concentric circles. LOS with no detections are shown as black crosses.}
\label{f:mach_ts_map}
\end{center}
\end{figure}

To verify if our fitting procedure is consistent with the results of previous analysis of our comparison sample, in Appendix~\ref{ap:sponge_comparison} we compare the distributions of the fitted Gaussian parameters (amplitude, line width, mean velocity) and derived physical properties ($N({\rm HI})_{\rm comp}$, $T_{K, \rm max}$, $T_s$, $\mathcal{M}_t$) for 21-SPONGE sources with the results of \citet{murray2018b}. We observe in Figure~\ref{f:sponge_comparison} that the results of the original processing of 21-SPONGE spectra are fully consistent with the method presented here within uncertainties. The results of all fits to the comparison sample are displayed in Appendix~\ref{ap:comparison_fits}.

\section{Results}
\label{sec:results}

\subsection{Properties integrated along LOS}

In Figure~\ref{f:integrated_properties} we display histograms of \nhi, \rhi, and \fcnm\ for the 44 MACH LOS. Our targets sample a low-column density (\nhi$\leq4\times10^{20}\rm\,cm^{-2}$), CNM-poor environment (median \rhi$=1.0$, median \fcnm$=0.0$). The maximum column density correction factor that we detect is \rhi$=1.10\pm 0.08$ and the maximum CNM fraction is \fcnm$=0.24\pm 0.11$.  

To investigate how the MACH target region compares with other high-latitude environments, we display cumulative distribution functions (CDFs) of \nhi, \rhi, and \fcnm\ for MACH and the comparison sample in Figure~\ref{f:integrated_param_comparison}. These properties for the \nabs\ MACH LOS and the comparison sample (\nrest\ LOS) are also summarized in Appendix A (Table~\ref{t:integrated}). The uncertainty ranges for the CDFs are estimated by bootstrapping each sample with replacement over $10^4$ trials, and represent the $1^{\rm st}$ through $99^{\rm th}$ percentiles of limits of the resulting distributions. 

Clearly, the MACH sample traces gas with significantly different \nhi, \rhi, and \fcnm\ than the comparison sample (LOS with $|b|>10^{\circ}$). The MACH CDFs in Figure~\ref{f:integrated_param_comparison} are fully discrepant within $\pm 3\sigma$ from the comparison sample.

\subsection{Properties of individual \hi\ structures}

Given that the integrated properties of MACH LOS are significantly different than the comparison sample, we investigate how the properties of individual \hi\ structures inferred from the Gaussian fitting procedure (Section~\ref{sec:gauss_fits}) compare. In Figure~\ref{f:fitted_param_comparison}, we display CDFs of the fitted Gaussian parameters (amplitude, line width, mean velocity) and derived physical properties ($N({\rm HI})_{\rm comp}$, $T_{K, \rm max}$, $T_s$, $\mathcal{M}_t$) for all components detected along MACH and comparison sample LOS. As in Figure~\ref{f:integrated_param_comparison}, the uncertainty ranges for the CDFs are estimated by bootstrapping the samples, and represent the  $1^{\rm st}$ through $99^{\rm th}$ percentiles of each distribution.

From Figure~\ref{f:fitted_param_comparison}, we observe that the amplitudes of Gaussian-fitted components (panel a) are statistically indistinguishable within uncertainties between the MACH and comparison samples. For the component line widths (FWHM; panel b), MACH features statistically more LOS with $\delta v <2.5\rm\,km\,s^{-1}$. In addition, the MACH LOS are dominated by components with mean velocities (panel c) $v_{0}<0\rm\,km\,s^{-1}$ relative to the comparison LOS.

The structures along MACH LOS sample a narrower range of $N({\rm HI})_{\rm comp}$ (Equation~\ref{e:nhi_comp}; panel d) than the comparison sample. The prevalence of low-$\delta v$ components results in smaller $T_{K,\rm max}$ (Equation~\ref{e:tkmax_comp}; panel e) for MACH relative to the comparison sample in Figure~\ref{f:fitted_param_comparison}. We observe that the spin temperature distributions (panel f) are statistically similar for low-$T_s$: for example, below the maximum $T_s$ for MACH ($T_s=163\pm6\rm\,K$) the samples are statistically indistinguishable within uncertainties, however, an additional $\sim 15\%$ of the comparison sample components have $T_s>170\rm\,K$. Given the narrower widths and smaller maximum kinetic temperatures, the $\mathcal{M}_t$ for MACH structures (Equation~\ref{e:mt_comp}) is statistically smaller than for the comparison sample structures (panel g).

\subsection{Latitude dependence of \hi\ properties}

To further investigate how the MACH environment compares with the rest of the high-latitude sky, we test how the integrated properties and fitted component properties vary with Galactic environment, parameterized by Galactic latitude. 

In Figure~\ref{f:integrated_param_latitude} we plot \nhi, \rhi, and \fcnm\ for the MACH and comparison samples as a function of absolute latitude ($|b|$). We observe that there is a clear correlation between \nhi\ and $|b|$ (panel a), where higher latitudes have lower \nhi. The MACH sample occupies the lower-\nhi\ end of the trend, but follows the same trend as the comparison sample. As $|b|$ decreases, each emission/absorption measurement samples a longer path length through the Milky Way, and therefore the resulting \nhi\ will sample more \hi\ structures and tend to be larger. In Figure~\ref{f:nhisinb}, we plot $N({\rm HI})\sin|b|$ vs. $\sin|b|$ for the MACH and comparison LOS, finding fully consistent results with previous analysis \citep[$N({\rm HI}) = 3.84\times10^{20} csc|b|$][]{heiles1976, dickey1990}. 
The variation in $N({\rm HI})\sin|b|$ with latitude arises because \hi\ is not organized in plane-parallel layers \citep{knapp1975, dickey1990}. The MACH LOS sample the minimum column densities, as a result of the structure of the local ISM, which we will discuss further in Section~\ref{sec:discussion}. 

In panels (b) and (c) of Figure~\ref{f:integrated_param_latitude} we observe a very mild trend in \rhi\ and \fcnm\ with $|b|$ (Pearson r correlation coefficients $-0.40$ and $-0.44$ respectively for the MACH and comparison samples combined). We observe that the largest values of \rhi\ and \fcnm\ are all at the lowest $|b|$ where the LOS sample the longest path lengths. 

In Figure~\ref{f:fitted_param_latitude}, we plot the fitted properties of individual \hi\ structures as a function of $|b|$ (same quantities as Figure~\ref{f:fitted_param_comparison}). To check for the presence of a correlation, we compute the Pearson r correlation coefficient for the full samples (MACH plus comparison) over $10^4$ trials, resampling in each trial. Here we use block bootstrapping, which involves breaking the sky into 10 equally-spaced bins (``blocks") in both longitude and latitude and resampling these blocks with replacement to incorporate the influence of large-scale interstellar structures into the parameter uncertainty. 

In all cases except $\tau_0$, $v_0$, and $N({\rm HI})_{\rm comp}$ the resulting Pearson r coefficient distributions are consistent with zero, indicating no significant correlation with $|b|$. For $\tau_0$, $v_0$, and $N({\rm HI})_{\rm comp}$, we observe negative correlation with $|b|$, significant at $>3\sigma$ (in terms of the block-bootstrapped distributions of the Pearson r coefficient).

The negative correlation between $v_0$ and $|b|$ arises because of the large-scale \hi\ distribution. For example, there are many well-known intermediate-velocity clouds (IVCs) in the Northern hemisphere, including the IV Arch and Spur \citep{kuntz1996} which skew the distribution of $v_0$ to negative velocities. We find that the majority of CNM components have central velocities $|v_0|<20\rm\,km\,s^{-1}$.  

To test for the effects of optical depth sensitivity on the correlation of $\tau_0$ and $N({\rm HI})_{\rm comp}$ with $|b|$, we repeat the correlation coefficient estimation for components binned by the RMS noise in the \thi\ spectrum they were fitted to ($\sigma_{\tau_{\rm HI}}$). In Figure~\ref{f:r_sensitivity} we plot the $50^{\rm th}$ percentile of the Pearson r coefficient ($|r|$) as a function of the maximum $3\sigma_{\tau_{\rm HI}}$ limit (i.e., for each bin, only components with $\sigma_{\tau_{\rm HI}}<3\times$ the indicated limit are included). We also split the samples between the Northern and Southern hemispheres. For both $\tau_0$ and $N({\rm HI})_{\rm comp}$, the negative correlation with $|b|$ persists across all sensitivity limits within block-bootstrapped uncertainties, for the full, Northern and Southern samples.

The negative correlation between $\tau_0$ and $|b|$ in Figure~\ref{f:fitted_param_latitude} (i.e., higher optical depths at lower latitudes) likely arises due to crowding effects. At lower latitudes where the spectral line complexity is higher, the completeness of Gaussian decomposition for recovering real individual \hi\ structures declines \citep{murray2017}. As a result, spectral features which are caused by multiple, blended, low-$\tau_0$ components may be incorrectly fit by fewer, higher-$\tau_0$ components. We selected our sample to have $|b|>10^{\circ}$ to reduce this bias, but it is still present. The correlation in $\tau_0$ propagates to drive the observed correlation with $N({\rm HI})_{\rm comp}$.

\begin{figure}
\begin{center}
\vspace{-20pt}
\includegraphics[width=0.47\textwidth]{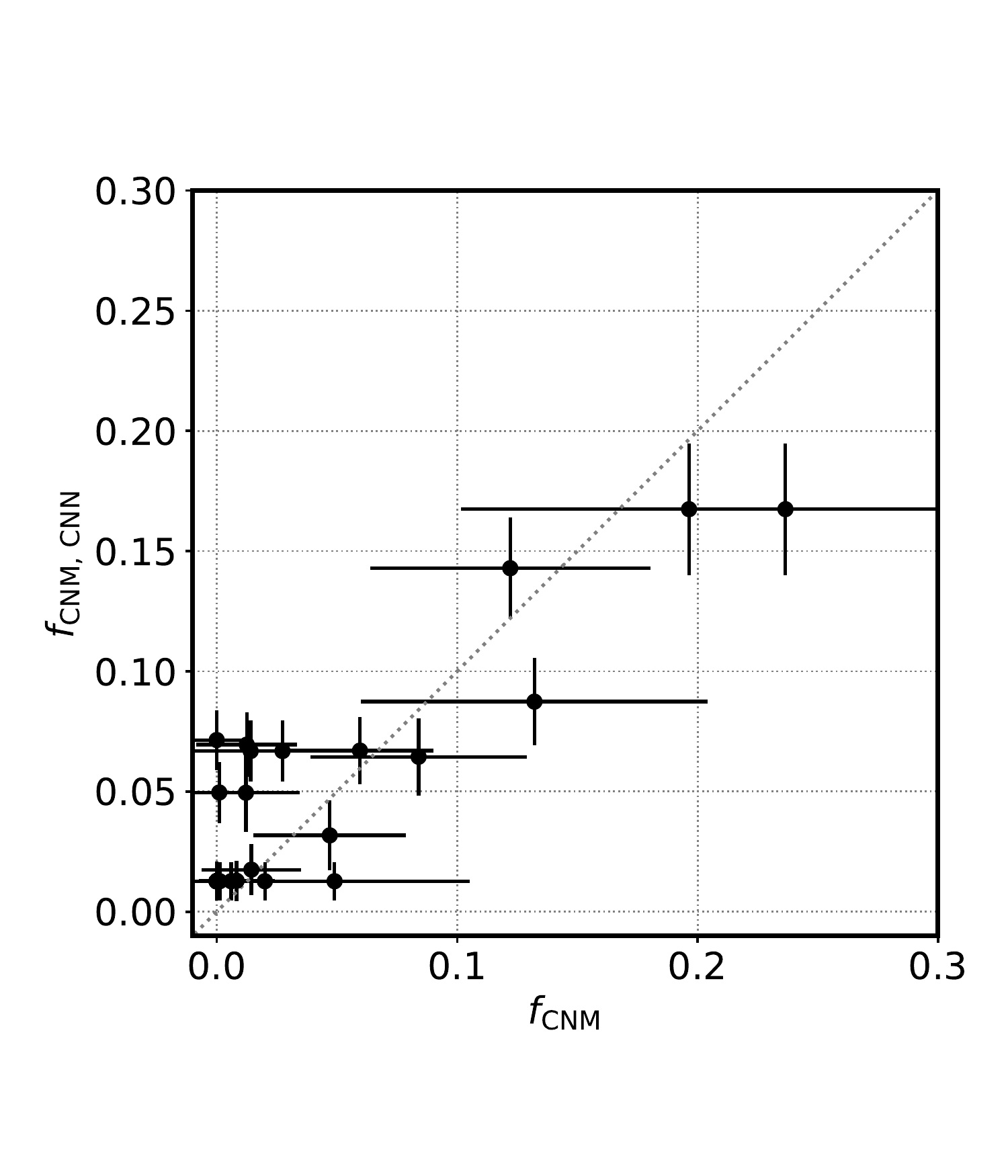}
\vspace{-40pt}
\caption{Comparing predictions of \fcnm\ from a convolutional neural network (\fcnm$_{,\rm CNN}$) applied to EBHIS with constraints from MACH (\fcnm). Within uncertainties, the \fcnm\ and \fcnm$_{,\rm CNN}$ are consistent.}
\label{f:fcnm_cnn}
\end{center}
\end{figure}

\begin{figure}
\begin{center}
\includegraphics[width=0.5\textwidth]{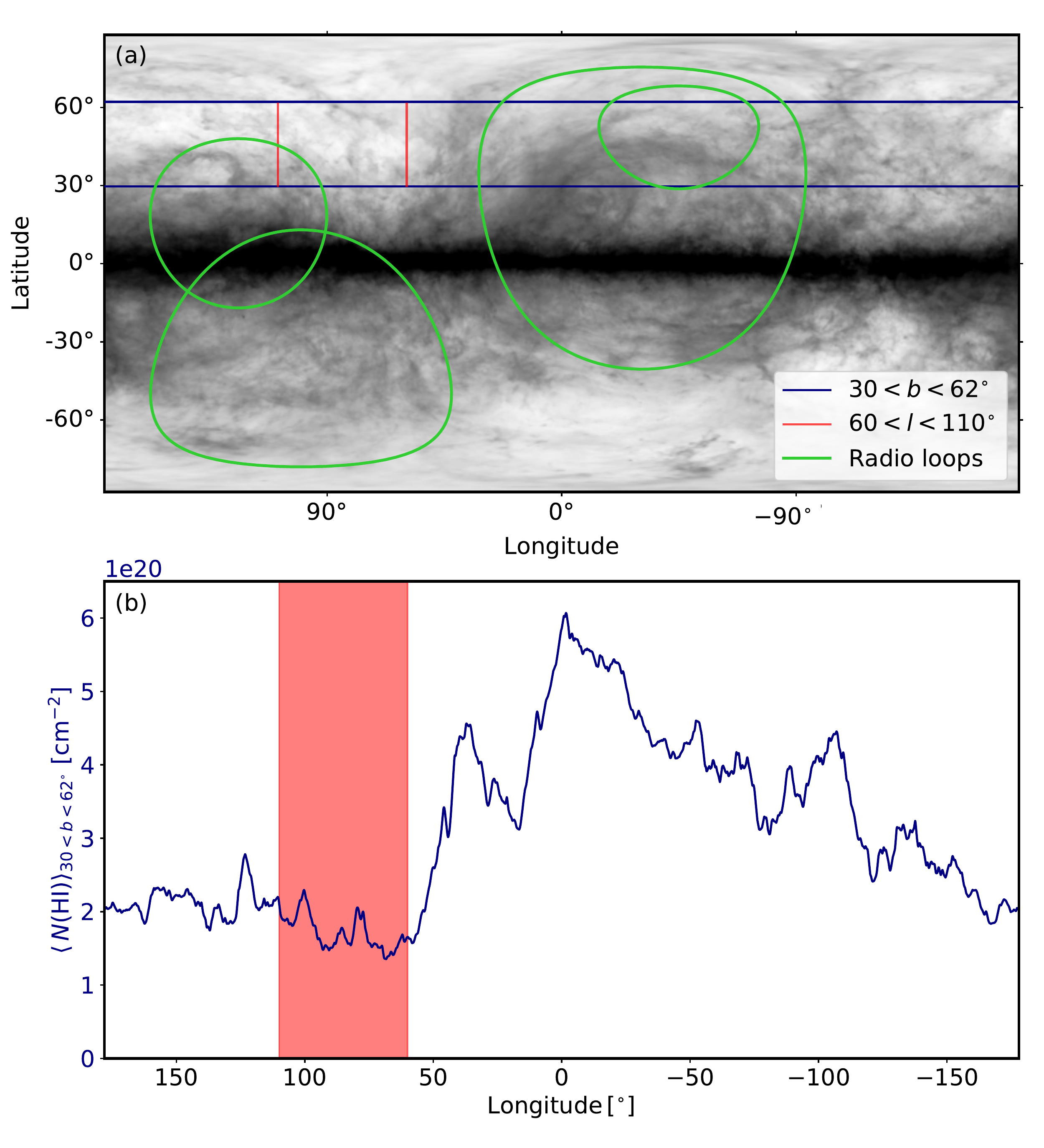}
\caption{(a): All-sky $N({\rm HI})^*$ map \citep[][; same scale as Figure~\ref{f:position_comparison}]{hi4pi} with latitude (red, vertical) and longitude (blue, horizontal) ranges for the MACH survey overlaid, including the outlines of known radio loops \citep{berkhuijsen1971}. (b): Mean $N({\rm HI})^*$ from $30<b<62^{\circ}$ as a function of longitude. The MACH survey (shaded red) covers the minimum column density for this latitude range.  }
\label{f:nhi_carl}
\end{center}
\end{figure}

\section{Discussion}
\label{sec:discussion}

Overall, we observe that the MACH sample probes a region with significantly low $N(\rm HI)$ and small \fcnm\ relative to the rest of the high and intermediate-latitude sky traced by previous \hi\ absorption surveys (Figure~\ref{f:integrated_param_comparison}). Although the properties of individual structures are generally similar as those detected across $|b|>10^{\circ}$ (Figure~\ref{f:fitted_param_comparison}), MACH LOS feature \hi\ structures with smaller $\delta v$, $T_{\rm K,max}$ and $\mathcal{M}_t$. These results were not expected when the survey was conceived, and below we discuss the implications. 

\begin{figure*}
\begin{center}
\includegraphics[width=0.9\textwidth]{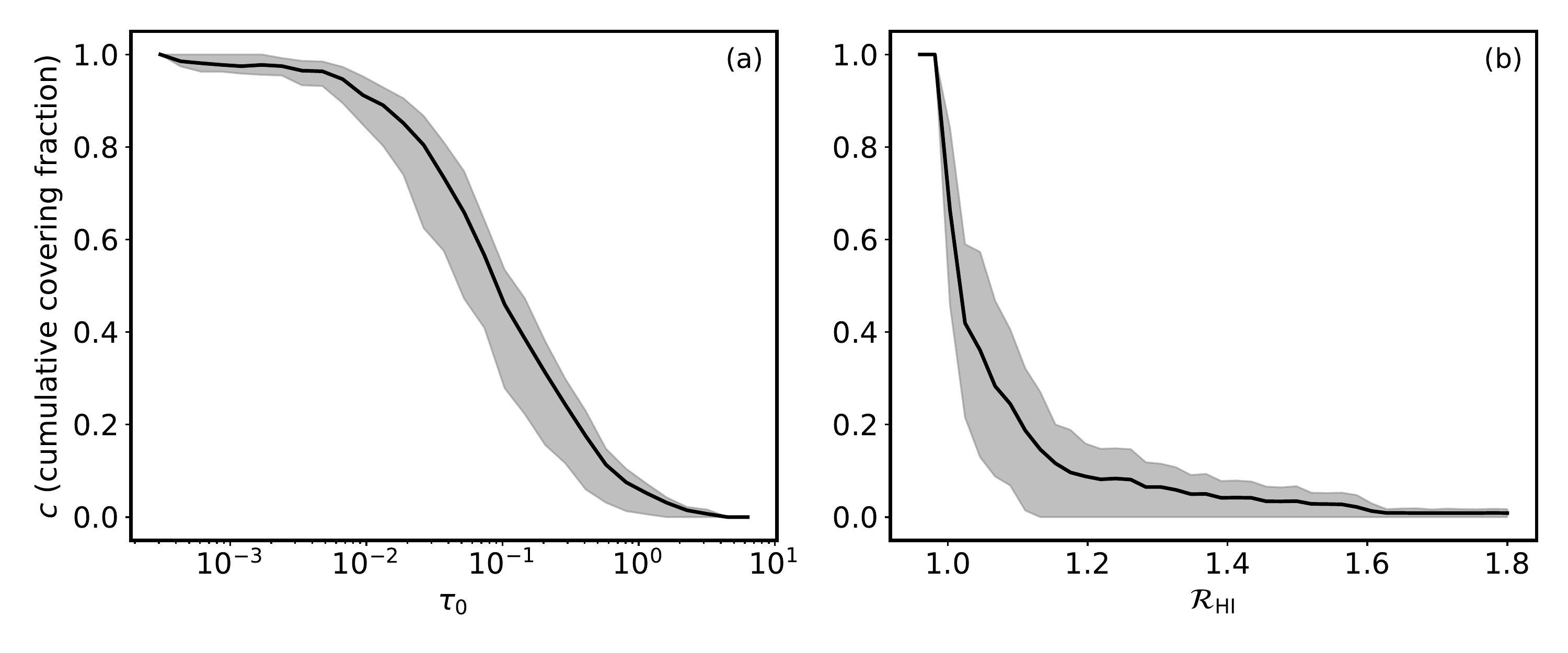}
\caption{Cumulative sky covering fractions as a function of $\tau_{\rm HI}$ (a) and \rhi\ (b) from the combined MACH and comparison samples. For (a), for each value of $\tau_0$, only sightlines with $\sigma_{\tau_{\rm HI}}\leq3\times\tau_0$ are included in the calculation. The uncertainties (shading) represent the $16^{\rm th}$ through $84^{\rm th}$ percentiles and are computed by block bootstrapping (see text). }
\label{f:covfrac}
\end{center}
\end{figure*}

That such a large region of the Northern sky could be so low-column and free of CNM is a reflection of how the local \hi\ distribution departs significantly from pure plane-parallel symmetry. It is well-known that the sun sits in a ``Local Bubble” or cavity, full of low-density gas \citep{cox1987}, surrounded by other bubble structures believed to be caused by star formation activity \citep[e.g.,][]{berkhuijsen1971}. The MACH field sits in a location where the bubble appears to be breaking out of the disk in a patchy, chimney-like structure \citep{lallement2003, vergely2010}, and the distance to the cool, absorbing \hi\ along MACH LOS is likely $150$ to $400\rm\,pc$, possibly as close as the edge of the bubble \citep{lallement2019}.

In Figure~\ref{f:mach_ts_map} we zoom-in on the MACH target region, and plot the LOS by their component $T_s$ estimates, with crosses for non-detections. This illustration emphasizes the patchy nature of the CNM distribution at high latitude -- for a roughly 400-square degree patch of sky ($60\lesssim l \lesssim100^{\circ}$, $40\lesssim b \lesssim50$) we detect no CNM at our optical depth sensitivity limit of $\sim 0.005$ per $0.42\rm\,km\,s^{-1}$ channels, and most detections are clustered together at the lowest latitudes. 

Could we have predicted that the MACH region would end up being so low-column and CNM-poor? To address this, we return to an inspection of Figure~\ref{f:spectra_summary}, where we observe that the structure of \hi\ emission on small scales near MACH LOS provides a reasonable prediction for the presence or absence of absorption lines. Channels featuring the strongest variation in $T_B(v)$ between off positions (quantified by $\sigma_{T_B}(v)$) tend to correspond with channels of detected absorption at our sensitivity \citep{dempsey2020}. This is not always the case -- for example, for source J14434 we detect absorption associated with only one of two narrow \hi\ emission features, which reflects the small-scale structure of the CNM in this region. In general, absorption features correspond with the narrowest velocity structures in $T_B(v)$, and LOS without detected absorption are free of discernible narrow velocity structure in $T_B(v)$. In agreement, \citet{murray2020} recently showed that a simple convolutional neural network (CNN) trained using synthetic \hi\ observations can predict \fcnm\ from the velocity structure of $T_B(v)$ alone (i.e., without \thi\ information). We applied their CNN model to MACH LOS and compare the resulting estimates (\fcnm$_{,CNN}$) with the constraints from absorption in Figure~\ref{f:fcnm_cnn}. Within uncertainties, the simple CNN accurately recovers the observed \fcnm\ values. In future studies, this or similar methods may be used to predict where \hi\ absorption is likely to be detected.

To investigate the state of the MACH region relative to similar high-latitude regions, we plot an all-sky \hi\ column density ($N({\rm HI})^*$) map in Figure~\ref{f:nhi_carl}a, highlighting the MACH latitude and longitude ranges. In Figure~\ref{f:nhi_carl}b we plot the mean $N({\rm HI})^*$ across MACH latitudes as a function of longitude, and observe that the MACH region has the lowest $N({\rm HI})^*$ for this latitude range. The high-latitude sky is populated by well-known loop structures (overlaid on Figure~\ref{f:nhi_carl}a), which likely trace shocked, swept-up magnetic fields and relativistic electrons, as well as swept-up \hi\ in fibrous structures from recent supernova activity \citep{berkhuijsen1971}.

That the morphology of the MACH region appears different (Figure~\ref{f:mach_ts_map}) from the swept-up shock picture suggests that the MACH region may not have been hit by a supernova shock recently. This scenario would be consistent with the finding in Figure~\ref{f:fitted_param_latitude} that CNM components detected along MACH LOS have significantly smaller turbulent MACH number (panel g) than the rest of the high and intermediate-latitude sky. In the relative absence of non-thermal motions from turbulence driven by star formation activity, the CNM line widths will be dominated by thermal motions alone, and $T_{k,\rm max}\approx T_s$. So, the MACH LOS may simply be tracing a particularly quiescent, undisturbed patch of sky. 

\subsection{The covering fraction of cold \hi}

Given the patchy nature of the CNM at high Galactic latitudes, we use the assembled sample of absorption constraints (the MACH and comparison samples combined) to estimate the covering fraction of cold \hi. 

In Figure~\ref{f:covfrac} we plot the cumulative covering fraction of cold \hi\ as a function of component optical depth ($\tau_0$; a) and column density correction factor (\rhi). The covering fractions, $c$ are computed as the fraction of all LOS featuring a component with optical depth greater than or equal to the indicated value (a), or a correction factor greater than or equal to the indicated value (b). In the case of $\tau_0$, we take into account the effects of varying optical depth sensitivity between LOS by only considering LOS with $\sigma_{\tau_{\rm HI}}\le 3 \tau_0$ for each bin. The uncertainties on the covering fractions are estimated by block-bootstrapping the samples, as discussed previously. 
We observe from Figure~\ref{f:covfrac}a that for \hi\ structures with $\tau_0>0.001$ $c\sim100\%$: at our sensitivity and within uncertainties, we cannot rule out the presence of this population of components along all LOS. For $\tau_0>0.01$, and $\tau_0>0.1$, $c\sim90\%$ and $c\sim 45\%$ respectively. By $\tau_0>1$, $c$ consistent with zero, suggesting that individual structures with $\tau_0>1$ are not present at high Galactic latitude ($|b|>10^{\circ}$).  

A consistent story is presented in Figure~\ref{f:covfrac}b. We observe that for LOS with \rhi$\geq1.003$, $c=65\%$, and for \rhi$\geq 1.2$ it is only $\sim10\%$. This further emphasizes the picture that the CNM at high Galactic latitude does not contribute significantly to the total \hi\ column density. As concluded by \citet{murray2018a} using a similar comparison sample from the literature, the lack of large \hi\ optical depths detected in absorption rules out the hypothesis that cold \hi\ is the dominant form of ``dark gas" at high latitude. 

However, despite the generally small optical depth of the CNM at high latitude, it is still a dynamically and physically-relevant phase in the ISM energy budget in this environment. Linear \hi\ structures (or fibers) observed in \hi\ emission which trace the local magnetic field \citep{clark2014, clark2015} are dominated by the cold \hi\ \citep{clark2019, peek2019b}. In addition, the correlation between \hi\ and dust emission varies with \hi\ column  density \citep{lenz2017, nguyen2018}, even within the low-column density environment probed here \citep{murray2020}, indicating that \hi\ phase balance influences the mixture between gas and dust -- critical for precise Galactic foreground estimation. 
 
\section{Summary and Conclusions}
\label{sec:summary}

We have presented the results of MACH, a survey of $21\rm\,cm$ absorption in a high-Galactic latitude patch of sky ($60<l<110^{\circ}$, $30<b<62^{\circ}$) with the VLA. We reach sufficient sensitivity in optical depth to detect absorption by the CNM in this region. For all \nabs\ LOS, we compute the total column density, fraction of CNM, and the column density correction due to optical depth, as well as the properties of individual \hi\ structures along each sightline via autonomous Gaussian decomposition. To compare the MACH region with the rest of the high and intermediate latitude sky ($|b|>10^{\circ}$), we analyze a sample of \nrest\ LOS from the literature and compare the results. Our main results are summarized as follows:

\begin{enumerate}
    \item{The integrated MACH LOS reveal surprisingly low-column density, low-\fcnm, low-\rhi\ gas. We find median \nhi$=1.5\times10^{20}\rm\,cm^{-2}$, \fcnm$=0.0$, \rhi$=1.0$ (Figure~\ref{f:integrated_properties}). Relative to the high- and intermediate-latitude sky probed by the comparison sample, these properties are significantly different (Figure~\ref{f:integrated_param_comparison}).}
    \item{Individual \hi\ structures along MACH LOS have generally similar properties to those detected along comparison sample LOS (Figure~\ref{f:fitted_param_comparison}). However, the line widths, kinetic temperatures and turbulent Mach numbers of MACH structures tend to be smaller. Although star formation activity has disrupted the local \hi\ structure in the high-latitude sky, producing a patchy distribution of CNM, the MACH LOS may be sampling a particularly low-column, quiescent region undisturbed recently by supernova shocks, leading its \hi\ properties to be dominated by thermal motions rather than non-thermal, turbulent motions.}
    \item{For the full sample (MACH and comparison combined) we compute the cumulative covering fraction of CNM properties at $|b|>10^{\circ}$ (Figure~\ref{f:covfrac}), and find that the high- and intermediate-latitude sky is dominated by gas with small optical depth and column density. For \hi\ structures with amplitude in optical depth of $\tau_0>0.001$, the covering fraction is consistent with $100\%$, For $\tau_0>0.01$, $\tau_0>0.1$ and $\tau_0>1$, the covering fractions are $c=90\%$, $45\%$ and $0\%$ respectively. In terms of the cumulative correction for optical depth (\rhi), for \rhi$\geq1.003$ the covering fraction is $c=65\%$ and \rhi$\geq 1.2$ it is only $10\%$. }
\end{enumerate}

Overall, the MACH LOS demonstrate the power of targeted $21\rm\,cm$ absorption studies to reveal unique interstellar environments in the local ISM. Future, expanded samples of Galactic absorption with similarly high optical depth sensitivity covering significantly larger fractions of sky are incoming with next-generation survey telescopes, including the Australian Square Kilometer Array Pathfinder \citep{dickey2013}. These samples will enable precise characterization of the CNM properties throughout the Milky Way. 

In the meantime, as future work we look forward to tackling additional science goals with the available samples. For example, the relative lack of blended, overlapping lines along MACH LOS make it a powerful sample for constraining the number of absorption lines as a function of optical depth, a calculation typically confounded by the masking of strong absorption lines. In addition, the MACH non-detection LOS provide lower limits to the spin temperature of gas without detected absorption, and can also impose limits on the minimum brightness temperature required for detecting absorption. Finally, we can compare predictions from 3D models of the local dust distribution based on IR emission, stellar photometry, and/or gamma rays \citep[e.g.,][]{remy2017, green2019, lallement2019, leike2020} with CNM inferred from $21\rm\,cm$ absorption to see if they predict consistent ISM structures in the $\sim100-500\rm\,pc$ distance regime. 

\facilities{VLA, Arecibo, Effelsberg}

\software{Astropy \citep{astropy2013}, NumPy \citep{vanderwalt2011}, matplotlib \citep{hunter2007}, GaussPy \citep{lindner2015}}

\acknowledgements{
We thank the referee for their thoughtful report which has improved this work. 
C.E.M. is supported by an NSF Astronomy and Astrophysics Postdoctoral Fellowship under award AST-1801471.
This work makes use of data from the Karl G. Jansky Very Large Array, operated by the National Radio Astronomy Observatory (NRAO). NRAO is a facility of the NSF operated under cooperative agreement by Associated Universities, Inc. This work also makes use of data from the Arecibo Observatory, which was operated by SRI International under a cooperative agreement with the National Science Foundation (AST-1100968), and in alliance with Ana G. M\'{e}ndez-Universidad Metropolitana, and the Universities Space Research Association. This work is partly based on observations with the 100-m telescope of the MPIfR (Max-Planck-Institut für Radioastronomie) at Effelsberg.
This research has made use of NASA's Astrophysics Data System. }

\appendix

\section{Integrated Properties}
\label{ap:int_properties}

In the following Appendix, we include a table of integrated properties for the MACH and comparison samples (Table~\ref{t:integrated}).

\startlongtable
\begin{deluxetable*}{llcc|cccc} 
\tablecaption{\label{t:integrated} Integrated Properties}
\tablehead{
\colhead{Name} & \colhead{Survey} & \colhead{l}    & \colhead{b} & \colhead{$N({\rm HI})_{\rm thin}$} & \colhead{$N({\rm HI})$} & \colhead{$\mathcal{R}_{\rm HI}$} & \colhead{$f_{\rm CNM}$}    \\ 
\colhead{} & \colhead{} & \colhead{($^{\circ}$)} & \colhead{($^{\circ}$)} & \colhead{($10^{20}\rm\,cm^{-2}$)} & \colhead{($10^{20}\rm\,cm^{-2}$)}  & \colhead{}  & \colhead{} \\ 
\colhead{(1)} & \colhead{(2)} & \colhead{(3)} & \colhead{(4)} & \colhead{(5)} & \colhead{(6)}  & \colhead{(7)} & \colhead{(8)} }
\startdata
J14002  &  MACH  &  109.589  &  53.127  &  1.40$\, \pm \,  $0.11  &  1.40$\, \pm \,  $0.11  &  1.00$\, \pm \,  $0.11  &  0.00$\, \pm \,  $0.00  \\ 
J14364  &  MACH  &  105.174  &  49.730  &  1.30$\, \pm \,  $0.08  &  1.30$\, \pm \,  $0.08  &  1.00$\, \pm \,  $0.09  &  0.00$\, \pm \,  $0.00  \\ 
J14384  &  MACH  &  103.524  &  50.695  &  1.50$\, \pm \,  $0.13  &  1.50$\, \pm \,  $0.13  &  1.00$\, \pm \,  $0.12  &  0.00$\, \pm \,  $0.00  \\ 
J14434  &  MACH  &  88.257  &  58.314  &  2.29$\, \pm \,  $0.23  &  2.29$\, \pm \,  $0.23  &  1.00$\, \pm \,  $0.14  &  0.01$\, \pm \,  $0.02  \\ 
J14510  &  MACH  &  71.913  &  61.477  &  1.46$\, \pm \,  $0.05  &  1.47$\, \pm \,  $0.05  &  1.00$\, \pm \,  $0.05  &  0.00$\, \pm \,  $0.01  \\ 
J14540  &  MACH  &  85.672  &  57.249  &  1.35$\, \pm \,  $0.04  &  1.35$\, \pm \,  $0.04  &  1.00$\, \pm \,  $0.04  &  0.00$\, \pm \,  $0.00  \\ 
J15040  &  MACH  &  97.691  &  50.106  &  1.11$\, \pm \,  $0.09  &  1.11$\, \pm \,  $0.09  &  1.00$\, \pm \,  $0.11  &  0.00$\, \pm \,  $0.00  \\ 
J15075  &  MACH  &  99.785  &  48.310  &  1.32$\, \pm \,  $0.09  &  1.32$\, \pm \,  $0.09  &  1.00$\, \pm \,  $0.10  &  0.00$\, \pm \,  $0.00  \\ 
J15102  &  MACH  &  87.319  &  53.685  &  1.70$\, \pm \,  $0.07  &  1.70$\, \pm \,  $0.07  &  1.00$\, \pm \,  $0.06  &  0.00$\, \pm \,  $0.00  \\ 
J15394  &  MACH  &  95.340  &  45.854  &  1.66$\, \pm \,  $0.11  &  1.66$\, \pm \,  $0.11  &  1.00$\, \pm \,  $0.09  &  0.00$\, \pm \,  $0.00  \\ 
J15412  &  MACH  &  61.493  &  52.905  &  1.18$\, \pm \,  $0.05  &  1.19$\, \pm \,  $0.05  &  1.00$\, \pm \,  $0.06  &  0.00$\, \pm \,  $0.01  \\ 
J15593  &  MACH  &  76.443  &  50.514  &  1.32$\, \pm \,  $0.07  &  1.32$\, \pm \,  $0.07  &  1.00$\, \pm \,  $0.07  &  0.00$\, \pm \,  $0.01  \\ 
J15452  &  MACH  &  74.158  &  50.850  &  1.25$\, \pm \,  $0.05  &  1.26$\, \pm \,  $0.05  &  1.00$\, \pm \,  $0.06  &  0.00$\, \pm \,  $0.02  \\ 
J15484  &  MACH  &  95.276  &  44.657  &  0.99$\, \pm \,  $0.10  &  0.99$\, \pm \,  $0.10  &  1.00$\, \pm \,  $0.15  &  0.00$\, \pm \,  $0.00  \\ 
J15512  &  MACH  &  97.835  &  43.242  &  1.41$\, \pm \,  $0.09  &  1.41$\, \pm \,  $0.09  &  1.00$\, \pm \,  $0.09  &  0.00$\, \pm \,  $0.01  \\ 
J16024  &  MACH  &  82.293  &  46.439  &  1.11$\, \pm \,  $0.06  &  1.11$\, \pm \,  $0.06  &  1.00$\, \pm \,  $0.07  &  0.00$\, \pm \,  $0.02  \\ 
J16042  &  MACH  &  92.926  &  43.385  &  1.06$\, \pm \,  $0.04  &  1.06$\, \pm \,  $0.04  &  1.00$\, \pm \,  $0.06  &  0.00$\, \pm \,  $0.01  \\ 
J16114  &  MACH  &  64.549  &  46.906  &  0.68$\, \pm \,  $0.07  &  0.69$\, \pm \,  $0.07  &  1.00$\, \pm \,  $0.15  &  0.05$\, \pm \,  $0.06  \\ 
J16255  &  MACH  &  65.744  &  44.219  &  0.78$\, \pm \,  $0.04  &  0.78$\, \pm \,  $0.04  &  1.00$\, \pm \,  $0.08  &  0.00$\, \pm \,  $0.00  \\ 
J16311  &  MACH  &  68.801  &  43.200  &  0.69$\, \pm \,  $0.04  &  0.69$\, \pm \,  $0.04  &  1.00$\, \pm \,  $0.07  &  0.00$\, \pm \,  $0.03  \\ 
J16343  &  MACH  &  93.611  &  39.384  &  1.97$\, \pm \,  $0.11  &  1.97$\, \pm \,  $0.11  &  1.00$\, \pm \,  $0.08  &  0.00$\, \pm \,  $0.00  \\ 
J16351  &  MACH  &  88.639  &  40.410  &  1.49$\, \pm \,  $0.18  &  1.49$\, \pm \,  $0.18  &  1.00$\, \pm \,  $0.17  &  0.00$\, \pm \,  $0.01  \\ 
J16403  &  MACH  &  61.602  &  41.327  &  1.02$\, \pm \,  $0.07  &  1.02$\, \pm \,  $0.07  &  1.00$\, \pm \,  $0.10  &  0.00$\, \pm \,  $0.00  \\ 
3C345  &  MACH  &  63.455  &  40.949  &  0.87$\, \pm \,  $0.04  &  0.87$\, \pm \,  $0.04  &  1.00$\, \pm \,  $0.07  &  0.00$\, \pm \,  $0.00  \\ 
J16480  &  MACH  &  60.857  &  39.799  &  1.88$\, \pm \,  $0.28  &  1.88$\, \pm \,  $0.28  &  1.00$\, \pm \,  $0.21  &  0.00$\, \pm \,  $0.00  \\ 
J16535  &  MACH  &  63.600  &  38.859  &  1.67$\, \pm \,  $0.11  &  1.67$\, \pm \,  $0.11  &  1.00$\, \pm \,  $0.09  &  0.00$\, \pm \,  $0.00  \\ 
J16572  &  MACH  &  85.740  &  37.857  &  1.83$\, \pm \,  $0.18  &  1.84$\, \pm \,  $0.18  &  1.00$\, \pm \,  $0.14  &  0.01$\, \pm \,  $0.02  \\ 
J16574  &  MACH  &  74.368  &  38.478  &  1.43$\, \pm \,  $0.07  &  1.44$\, \pm \,  $0.07  &  1.00$\, \pm \,  $0.07  &  0.01$\, \pm \,  $0.02  \\ 
J16580  &  MACH  &  73.714  &  38.439  &  1.66$\, \pm \,  $0.04  &  1.67$\, \pm \,  $0.04  &  1.01$\, \pm \,  $0.03  &  0.05$\, \pm \,  $0.03  \\ 
J16582  &  MACH  &  62.883  &  37.929  &  1.73$\, \pm \,  $0.09  &  1.73$\, \pm \,  $0.09  &  1.00$\, \pm \,  $0.08  &  0.00$\, \pm \,  $0.01  \\ 
J17024  &  MACH  &  83.347  &  37.317  &  1.36$\, \pm \,  $0.06  &  1.36$\, \pm \,  $0.06  &  1.00$\, \pm \,  $0.06  &  0.02$\, \pm \,  $0.03  \\ 
J17025  &  MACH  &  77.079  &  37.600  &  1.76$\, \pm \,  $0.14  &  1.77$\, \pm \,  $0.14  &  1.01$\, \pm \,  $0.11  &  0.00$\, \pm \,  $0.02  \\ 
J17054  &  MACH  &  79.516  &  37.094  &  2.21$\, \pm \,  $0.16  &  2.32$\, \pm \,  $0.17  &  1.05$\, \pm \,  $0.10  &  0.13$\, \pm \,  $0.07  \\ 
J17104A  &  MACH  &  71.763  &  36.230  &  2.31$\, \pm \,  $0.13  &  2.55$\, \pm \,  $0.14  &  1.10$\, \pm \,  $0.08  &  0.24$\, \pm \,  $0.11  \\ 
J17104B  &  MACH  &  71.763  &  36.227  &  2.31$\, \pm \,  $0.13  &  2.53$\, \pm \,  $0.14  &  1.09$\, \pm \,  $0.08  &  0.20$\, \pm \,  $0.09  \\ 
J17195  &  MACH  &  93.806  &  34.144  &  2.70$\, \pm \,  $0.09  &  2.70$\, \pm \,  $0.09  &  1.00$\, \pm \,  $0.05  &  0.00$\, \pm \,  $0.01  \\ 
J17233  &  MACH  &  79.919  &  34.350  &  2.68$\, \pm \,  $0.09  &  2.71$\, \pm \,  $0.09  &  1.01$\, \pm \,  $0.05  &  0.00$\, \pm \,  $0.01  \\ 
J17251B  &  MACH  &  65.568  &  33.018  &  2.73$\, \pm \,  $0.18  &  2.77$\, \pm \,  $0.18  &  1.02$\, \pm \,  $0.09  &  0.03$\, \pm \,  $0.03  \\ 
J17251A  &  MACH  &  65.569  &  33.016  &  2.73$\, \pm \,  $0.18  &  2.76$\, \pm \,  $0.18  &  1.01$\, \pm \,  $0.09  &  0.01$\, \pm \,  $0.03  \\ 
J17304  &  MACH  &  75.772  &  33.068  &  2.39$\, \pm \,  $0.15  &  2.50$\, \pm \,  $0.15  &  1.04$\, \pm \,  $0.09  &  0.12$\, \pm \,  $0.06  \\ 
J17305  &  MACH  &  62.989  &  31.515  &  3.28$\, \pm \,  $0.11  &  3.34$\, \pm \,  $0.11  &  1.02$\, \pm \,  $0.05  &  0.08$\, \pm \,  $0.05  \\ 
J17395  &  MACH  &  74.222  &  31.396  &  2.08$\, \pm \,  $0.12  &  2.10$\, \pm \,  $0.12  &  1.01$\, \pm \,  $0.08  &  0.01$\, \pm \,  $0.02  \\ 
J17403  &  MACH  &  79.563  &  31.748  &  2.89$\, \pm \,  $0.07  &  2.93$\, \pm \,  $0.07  &  1.02$\, \pm \,  $0.03  &  0.01$\, \pm \,  $0.02  \\ 
J17422  &  MACH  &  81.770  &  31.613  &  3.12$\, \pm \,  $0.10  &  3.16$\, \pm \,  $0.10  &  1.01$\, \pm \,  $0.05  &  0.06$\, \pm \,  $0.03  \\ 
3C327.1A  &  21-SPONGE  &  12.181  &  37.006  &  6.98$\, \pm \,  $0.24  &  7.67$\, \pm \,  $0.28  &  1.10$\, \pm \,  $0.05  &  0.25$\, \pm \,  $0.09  \\ 
4C16.09  &  21-SPONGE  &  166.636  &  -33.596  &  9.55$\, \pm \,  $0.36  &  10.56$\, \pm \,  $0.43  &  1.11$\, \pm \,  $0.06  &  0.23$\, \pm \,  $0.09  \\ 
1055+018  &  21-SPONGE  &  251.511  &  52.774  &  2.85$\, \pm \,  $0.09  &  2.85$\, \pm \,  $0.09  &  1.00$\, \pm \,  $0.05  &  0.00$\, \pm \,  $0.01  \\ 
3C459  &  21-SPONGE  &  83.040  &  -51.285  &  5.27$\, \pm \,  $0.13  &  5.43$\, \pm \,  $0.14  &  1.03$\, \pm \,  $0.04  &  0.17$\, \pm \,  $0.06  \\ 
3C018A  &  21-SPONGE  &  118.623  &  -52.732  &  5.72$\, \pm \,  $0.15  &  6.41$\, \pm \,  $0.18  &  1.12$\, \pm \,  $0.04  &  0.32$\, \pm \,  $0.12  \\ 
PKS2127  &  21-SPONGE  &  58.652  &  -31.815  &  4.31$\, \pm \,  $0.21  &  4.40$\, \pm \,  $0.22  &  1.02$\, \pm \,  $0.07  &  0.09$\, \pm \,  $0.04  \\ 
3C286  &  21-SPONGE  &  56.524  &  80.675  &  1.07$\, \pm \,  $0.04  &  1.07$\, \pm \,  $0.04  &  1.00$\, \pm \,  $0.06  &  0.03$\, \pm \,  $0.02  \\ 
3C298  &  21-SPONGE  &  352.160  &  60.666  &  1.87$\, \pm \,  $0.04  &  1.87$\, \pm \,  $0.04  &  1.00$\, \pm \,  $0.03  &  0.01$\, \pm \,  $0.01  \\ 
3C245B  &  21-SPONGE  &  233.123  &  56.299  &  2.30$\, \pm \,  $0.08  &  2.31$\, \pm \,  $0.08  &  1.00$\, \pm \,  $0.05  &  0.02$\, \pm \,  $0.02  \\ 
3C123B  &  21-SPONGE  &  170.578  &  -11.659  &  14.69$\, \pm \,  $1.46  &  18.75$\, \pm \,  $2.20  &  1.28$\, \pm \,  $0.16  &  0.41$\, \pm \,  $0.16  \\ 
3C454.3  &  21-SPONGE  &  86.112  &  -38.185  &  6.85$\, \pm \,  $0.33  &  7.03$\, \pm \,  $0.33  &  1.03$\, \pm \,  $0.07  &  0.22$\, \pm \,  $0.08  \\ 
3C225B  &  21-SPONGE  &  220.011  &  44.009  &  3.48$\, \pm \,  $0.17  &  3.60$\, \pm \,  $0.16  &  1.03$\, \pm \,  $0.07  &  0.38$\, \pm \,  $0.14  \\ 
3C273  &  21-SPONGE  &  289.945  &  64.359  &  1.64$\, \pm \,  $0.09  &  1.64$\, \pm \,  $0.09  &  1.00$\, \pm \,  $0.08  &  0.01$\, \pm \,  $0.01  \\ 
UGC09799  &  21-SPONGE  &  9.417  &  50.120  &  2.58$\, \pm \,  $0.10  &  2.59$\, \pm \,  $0.10  &  1.01$\, \pm \,  $0.05  &  0.03$\, \pm \,  $0.02  \\ 
3C041B  &  21-SPONGE  &  131.374  &  -29.070  &  5.01$\, \pm \,  $0.12  &  5.05$\, \pm \,  $0.12  &  1.01$\, \pm \,  $0.03  &  0.04$\, \pm \,  $0.02  \\ 
J2232  &  21-SPONGE  &  77.438  &  -38.582  &  4.71$\, \pm \,  $0.24  &  4.81$\, \pm \,  $0.25  &  1.02$\, \pm \,  $0.07  &  0.18$\, \pm \,  $0.07  \\ 
3C236  &  21-SPONGE  &  190.065  &  53.980  &  0.76$\, \pm \,  $0.06  &  0.76$\, \pm \,  $0.06  &  1.00$\, \pm \,  $0.12  &  0.00$\, \pm \,  $0.00  \\ 
J0022  &  21-SPONGE  &  107.462  &  -61.748  &  2.53$\, \pm \,  $0.10  &  2.54$\, \pm \,  $0.10  &  1.00$\, \pm \,  $0.05  &  0.02$\, \pm \,  $0.02  \\ 
3C263.1  &  21-SPONGE  &  227.201  &  73.766  &  1.66$\, \pm \,  $0.10  &  1.66$\, \pm \,  $0.10  &  1.00$\, \pm \,  $0.09  &  0.01$\, \pm \,  $0.01  \\ 
J1613  &  21-SPONGE  &  55.151  &  46.379  &  1.29$\, \pm \,  $0.07  &  1.29$\, \pm \,  $0.07  &  1.00$\, \pm \,  $0.08  &  0.00$\, \pm \,  $0.00  \\ 
3C123A  &  21-SPONGE  &  170.584  &  -11.660  &  14.69$\, \pm \,  $1.46  &  18.84$\, \pm \,  $2.21  &  1.28$\, \pm \,  $0.16  &  0.42$\, \pm \,  $0.17  \\ 
3C245A  &  21-SPONGE  &  233.124  &  56.300  &  2.30$\, \pm \,  $0.08  &  2.31$\, \pm \,  $0.08  &  1.00$\, \pm \,  $0.05  &  0.00$\, \pm \,  $0.01  \\ 
3C48  &  21-SPONGE  &  133.963  &  -28.719  &  4.13$\, \pm \,  $0.11  &  4.16$\, \pm \,  $0.11  &  1.01$\, \pm \,  $0.04  &  0.06$\, \pm \,  $0.03  \\ 
PKS0742  &  21-SPONGE  &  209.797  &  16.592  &  2.77$\, \pm \,  $0.06  &  2.77$\, \pm \,  $0.06  &  1.00$\, \pm \,  $0.03  &  0.00$\, \pm \,  $0.00  \\ 
4C12.50  &  21-SPONGE  &  347.223  &  70.172  &  1.90$\, \pm \,  $0.07  &  1.93$\, \pm \,  $0.07  &  1.01$\, \pm \,  $0.05  &  0.12$\, \pm \,  $0.05  \\ 
3C018B  &  21-SPONGE  &  118.616  &  -52.719  &  5.71$\, \pm \,  $0.17  &  6.34$\, \pm \,  $0.20  &  1.11$\, \pm \,  $0.04  &  0.31$\, \pm \,  $0.11  \\ 
3C327.1B  &  21-SPONGE  &  12.182  &  37.003  &  6.98$\, \pm \,  $0.24  &  7.63$\, \pm \,  $0.28  &  1.09$\, \pm \,  $0.05  &  0.23$\, \pm \,  $0.09  \\ 
3C132  &  21-SPONGE  &  178.862  &  -12.522  &  21.33$\, \pm \,  $0.26  &  25.25$\, \pm \,  $0.27  &  1.18$\, \pm \,  $0.02  &  0.23$\, \pm \,  $0.09  \\ 
4C32.44  &  21-SPONGE  &  67.234  &  81.048  &  1.21$\, \pm \,  $0.08  &  1.21$\, \pm \,  $0.08  &  1.00$\, \pm \,  $0.10  &  0.02$\, \pm \,  $0.02  \\ 
3C120  &  21-SPONGE  &  190.373  &  -27.397  &  10.39$\, \pm \,  $0.33  &  16.39$\, \pm \,  $0.48  &  1.58$\, \pm \,  $0.04  &  0.58$\, \pm \,  $0.21  \\ 
4C04.51  &  21-SPONGE  &  7.292  &  47.747  &  3.61$\, \pm \,  $0.11  &  3.65$\, \pm \,  $0.11  &  1.01$\, \pm \,  $0.04  &  0.05$\, \pm \,  $0.03  \\ 
J2136  &  21-SPONGE  &  55.473  &  -35.578  &  4.30$\, \pm \,  $0.34  &  4.42$\, \pm \,  $0.35  &  1.03$\, \pm \,  $0.11  &  0.16$\, \pm \,  $0.06  \\ 
4C25.43  &  21-SPONGE  &  22.468  &  80.988  &  0.88$\, \pm \,  $0.07  &  0.88$\, \pm \,  $0.07  &  1.00$\, \pm \,  $0.11  &  0.00$\, \pm \,  $0.00  \\ 
4C15.05  &  21-SPONGE  &  147.930  &  -44.043  &  4.63$\, \pm \,  $0.22  &  4.73$\, \pm \,  $0.22  &  1.02$\, \pm \,  $0.07  &  0.11$\, \pm \,  $0.05  \\ 
3C041A  &  21-SPONGE  &  131.379  &  -29.075  &  5.01$\, \pm \,  $0.12  &  5.05$\, \pm \,  $0.12  &  1.01$\, \pm \,  $0.03  &  0.04$\, \pm \,  $0.02  \\ 
3C78  &  21-SPONGE  &  174.858  &  -44.514  &  10.24$\, \pm \,  $0.17  &  11.81$\, \pm \,  $0.18  &  1.15$\, \pm \,  $0.02  &  0.36$\, \pm \,  $0.13  \\ 
PKS1607  &  21-SPONGE  &  44.171  &  46.203  &  3.52$\, \pm \,  $0.19  &  3.61$\, \pm \,  $0.20  &  1.03$\, \pm \,  $0.08  &  0.21$\, \pm \,  $0.08  \\ 
3C147  &  21-SPONGE  &  161.686  &  10.298  &  18.38$\, \pm \,  $0.64  &  20.08$\, \pm \,  $0.71  &  1.09$\, \pm \,  $0.05  &  0.20$\, \pm \,  $0.08  \\ 
3C225A  &  21-SPONGE  &  220.010  &  44.008  &  3.48$\, \pm \,  $0.17  &  3.61$\, \pm \,  $0.16  &  1.03$\, \pm \,  $0.07  &  0.37$\, \pm \,  $0.13  \\ 
3C138  &  21-SPONGE  &  187.405  &  -11.343  &  20.28$\, \pm \,  $1.07  &  23.14$\, \pm \,  $1.21  &  1.14$\, \pm \,  $0.07  &  0.21$\, \pm \,  $0.08  \\ 
3C346  &  21-SPONGE  &  35.332  &  35.769  &  4.97$\, \pm \,  $0.11  &  5.16$\, \pm \,  $0.11  &  1.04$\, \pm \,  $0.03  &  0.19$\, \pm \,  $0.07  \\ 
3C237  &  21-SPONGE  &  232.117  &  46.627  &  1.67$\, \pm \,  $0.09  &  1.69$\, \pm \,  $0.09  &  1.01$\, \pm \,  $0.08  &  0.34$\, \pm \,  $0.13  \\ 
3C433  &  21-SPONGE  &  74.475  &  -17.697  &  8.09$\, \pm \,  $0.17  &  8.59$\, \pm \,  $0.19  &  1.06$\, \pm \,  $0.03  &  0.18$\, \pm \,  $0.07  \\ 
NV0232+34  &  Perseus  &  145.598  &  -23.984  &  5.44$\, \pm \,  $0.35  &  5.71$\, \pm \,  $0.37  &  1.05$\, \pm \,  $0.09  &  0.20$\, \pm \,  $0.08  \\ 
4C+26.12  &  Perseus  &  165.818  &  -21.058  &  6.61$\, \pm \,  $0.21  &  6.95$\, \pm \,  $0.22  &  1.05$\, \pm \,  $0.05  &  0.26$\, \pm \,  $0.10  \\ 
3C093.1  &  Perseus  &  160.037  &  -15.914  &  10.60$\, \pm \,  $0.64  &  14.64$\, \pm \,  $0.66  &  1.38$\, \pm \,  $0.08  &  0.49$\, \pm \,  $0.18  \\ 
3C067  &  Perseus  &  146.822  &  -30.696  &  7.58$\, \pm \,  $0.17  &  8.28$\, \pm \,  $0.19  &  1.09$\, \pm \,  $0.03  &  0.26$\, \pm \,  $0.10  \\ 
4C+27.07  &  Perseus  &  145.012  &  -31.093  &  6.41$\, \pm \,  $0.26  &  6.69$\, \pm \,  $0.28  &  1.04$\, \pm \,  $0.06  &  0.15$\, \pm \,  $0.06  \\ 
3C108  &  Perseus  &  171.872  &  -20.117  &  10.28$\, \pm \,  $0.28  &  11.86$\, \pm \,  $0.31  &  1.15$\, \pm \,  $0.04  &  0.33$\, \pm \,  $0.12  \\ 
B20218+35  &  Perseus  &  142.602  &  -23.487  &  6.22$\, \pm \,  $0.31  &  6.33$\, \pm \,  $0.31  &  1.02$\, \pm \,  $0.07  &  0.06$\, \pm \,  $0.03  \\ 
4C+25.14  &  Perseus  &  171.372  &  -17.162  &  9.73$\, \pm \,  $0.56  &  11.04$\, \pm \,  $0.59  &  1.13$\, \pm \,  $0.08  &  0.34$\, \pm \,  $0.12  \\ 
NV0157+28  &  Perseus  &  139.899  &  -31.835  &  5.63$\, \pm \,  $0.13  &  5.71$\, \pm \,  $0.13  &  1.01$\, \pm \,  $0.03  &  0.05$\, \pm \,  $0.03  \\ 
B20400+25  &  Perseus  &  168.026  &  -19.648  &  7.98$\, \pm \,  $0.44  &  8.45$\, \pm \,  $0.47  &  1.06$\, \pm \,  $0.08  &  0.17$\, \pm \,  $0.07  \\ 
3C092  &  Perseus  &  159.738  &  -18.407  &  11.64$\, \pm \,  $0.60  &  18.55$\, \pm \,  $0.76  &  1.59$\, \pm \,  $0.07  &  0.59$\, \pm \,  $0.21  \\ 
5C06.237  &  Perseus  &  143.882  &  -26.525  &  5.56$\, \pm \,  $0.12  &  5.96$\, \pm \,  $0.13  &  1.07$\, \pm \,  $0.03  &  0.21$\, \pm \,  $0.08  \\ 
4C+34.09  &  Perseus  &  150.935  &  -20.486  &  9.50$\, \pm \,  $0.29  &  10.26$\, \pm \,  $0.31  &  1.08$\, \pm \,  $0.04  &  0.25$\, \pm \,  $0.09  \\ 
4C+28.06  &  Perseus  &  148.781  &  -28.443  &  7.50$\, \pm \,  $0.21  &  8.13$\, \pm \,  $0.22  &  1.08$\, \pm \,  $0.04  &  0.26$\, \pm \,  $0.10  \\ 
4C+34.07  &  Perseus  &  144.312  &  -24.550  &  5.54$\, \pm \,  $0.12  &  5.88$\, \pm \,  $0.13  &  1.06$\, \pm \,  $0.03  &  0.21$\, \pm \,  $0.08  \\ 
4C+28.07  &  Perseus  &  149.466  &  -28.528  &  7.55$\, \pm \,  $0.11  &  8.43$\, \pm \,  $0.13  &  1.12$\, \pm \,  $0.02  &  0.34$\, \pm \,  $0.12  \\ 
B20326+27  &  Perseus  &  160.703  &  -23.074  &  10.06$\, \pm \,  $0.30  &  11.00$\, \pm \,  $0.33  &  1.09$\, \pm \,  $0.04  &  0.23$\, \pm \,  $0.08  \\ 
B20411+34  &  Perseus  &  163.798  &  -11.981  &  16.46$\, \pm \,  $0.18  &  18.86$\, \pm \,  $0.17  &  1.15$\, \pm \,  $0.01  &  0.25$\, \pm \,  $0.09  \\ 
3C068.2  &  Perseus  &  147.326  &  -26.377  &  7.57$\, \pm \,  $0.33  &  8.56$\, \pm \,  $0.37  &  1.13$\, \pm \,  $0.06  &  0.31$\, \pm \,  $0.12  \\ 
4C+30.04  &  Perseus  &  155.401  &  -23.171  &  10.46$\, \pm \,  $0.20  &  12.13$\, \pm \,  $0.23  &  1.16$\, \pm \,  $0.03  &  0.37$\, \pm \,  $0.14  \\ 
4C+32.14  &  Perseus  &  159.000  &  -18.765  &  12.25$\, \pm \,  $0.14  &  27.08$\, \pm \,  $0.37  &  2.21$\, \pm \,  $0.03  &  0.67$\, \pm \,  $0.23  \\ 
4C+29.05  &  Perseus  &  140.716  &  -30.875  &  4.78$\, \pm \,  $0.09  &  4.89$\, \pm \,  $0.09  &  1.02$\, \pm \,  $0.03  &  0.09$\, \pm \,  $0.04  \\ 
3C105  &  HT03  &  187.633  &  -33.609  &  10.57$\, \pm \,  $0.30  &  17.08$\, \pm \,  $0.43  &  1.62$\, \pm \,  $0.04  &  0.61$\, \pm \,  $0.22  \\ 
3C109  &  HT03  &  181.828  &  -27.777  &  14.99$\, \pm \,  $0.34  &  22.54$\, \pm \,  $0.32  &  1.50$\, \pm \,  $0.03  &  0.46$\, \pm \,  $0.16  \\ 
3C142.1  &  HT03  &  197.616  &  -14.512  &  19.05$\, \pm \,  $0.26  &  25.67$\, \pm \,  $0.39  &  1.35$\, \pm \,  $0.02  &  0.34$\, \pm \,  $0.12  \\ 
3C172.0  &  HT03  &  191.205  &  13.410  &  8.47$\, \pm \,  $0.17  &  8.52$\, \pm \,  $0.17  &  1.01$\, \pm \,  $0.03  &  0.00$\, \pm \,  $0.00  \\ 
3C190.0  &  HT03  &  207.624  &  21.841  &  3.22$\, \pm \,  $0.14  &  3.22$\, \pm \,  $0.14  &  1.00$\, \pm \,  $0.06  &  0.00$\, \pm \,  $0.00  \\ 
3C192  &  HT03  &  197.913  &  26.410  &  4.57$\, \pm \,  $0.27  &  4.61$\, \pm \,  $0.27  &  1.01$\, \pm \,  $0.08  &  0.07$\, \pm \,  $0.03  \\ 
3C207  &  HT03  &  212.968  &  30.139  &  5.46$\, \pm \,  $0.37  &  5.73$\, \pm \,  $0.38  &  1.05$\, \pm \,  $0.09  &  0.32$\, \pm \,  $0.12  \\ 
3C228.0  &  HT03  &  220.831  &  46.635  &  2.81$\, \pm \,  $0.06  &  2.84$\, \pm \,  $0.06  &  1.01$\, \pm \,  $0.03  &  0.08$\, \pm \,  $0.04  \\ 
3C234  &  HT03  &  200.205  &  52.705  &  1.54$\, \pm \,  $0.10  &  1.54$\, \pm \,  $0.10  &  1.00$\, \pm \,  $0.09  &  0.13$\, \pm \,  $0.10  \\ 
3C264.0  &  HT03  &  236.996  &  73.642  &  2.94$\, \pm \,  $0.31  &  2.94$\, \pm \,  $0.31  &  1.00$\, \pm \,  $0.15  &  0.00$\, \pm \,  $0.01  \\ 
3C267.0  &  HT03  &  256.342  &  70.110  &  2.56$\, \pm \,  $0.11  &  2.56$\, \pm \,  $0.11  &  1.00$\, \pm \,  $0.06  &  0.00$\, \pm \,  $0.02  \\ 
3C272.1  &  HT03  &  280.632  &  74.687  &  2.02$\, \pm \,  $0.15  &  2.02$\, \pm \,  $0.15  &  1.00$\, \pm \,  $0.10  &  0.01$\, \pm \,  $0.03  \\ 
3C274.1  &  HT03  &  269.873  &  83.164  &  2.36$\, \pm \,  $0.14  &  2.39$\, \pm \,  $0.14  &  1.01$\, \pm \,  $0.08  &  0.15$\, \pm \,  $0.06  \\ 
3C293  &  HT03  &  54.607  &  76.060  &  1.23$\, \pm \,  $0.08  &  1.23$\, \pm \,  $0.08  &  1.00$\, \pm \,  $0.10  &  0.00$\, \pm \,  $0.02  \\ 
3C310  &  HT03  &  38.500  &  60.210  &  3.54$\, \pm \,  $0.18  &  3.81$\, \pm \,  $0.20  &  1.08$\, \pm \,  $0.07  &  0.35$\, \pm \,  $0.13  \\ 
3C315  &  HT03  &  39.360  &  58.303  &  4.47$\, \pm \,  $0.06  &  5.03$\, \pm \,  $0.07  &  1.13$\, \pm \,  $0.02  &  0.43$\, \pm \,  $0.15  \\ 
3C33-1  &  HT03  &  129.439  &  -49.343  &  2.90$\, \pm \,  $0.10  &  2.93$\, \pm \,  $0.10  &  1.01$\, \pm \,  $0.05  &  0.02$\, \pm \,  $0.03  \\ 
3C33-2  &  HT03  &  129.462  &  -49.277  &  3.01$\, \pm \,  $0.17  &  3.06$\, \pm \,  $0.17  &  1.02$\, \pm \,  $0.08  &  0.10$\, \pm \,  $0.05  \\ 
3C33  &  HT03  &  129.448  &  -49.324  &  2.98$\, \pm \,  $0.15  &  3.00$\, \pm \,  $0.15  &  1.01$\, \pm \,  $0.07  &  0.03$\, \pm \,  $0.02  \\ 
3C348  &  HT03  &  22.971  &  29.177  &  5.72$\, \pm \,  $0.25  &  6.20$\, \pm \,  $0.27  &  1.08$\, \pm \,  $0.06  &  0.29$\, \pm \,  $0.11  \\ 
3C353  &  HT03  &  21.111  &  19.877  &  9.55$\, \pm \,  $0.83  &  12.53$\, \pm \,  $1.21  &  1.31$\, \pm \,  $0.13  &  0.45$\, \pm \,  $0.18  \\ 
3C454.0  &  HT03  &  88.100  &  -35.941  &  4.51$\, \pm \,  $0.16  &  4.57$\, \pm \,  $0.17  &  1.01$\, \pm \,  $0.05  &  0.10$\, \pm \,  $0.04  \\ 
3C64  &  HT03  &  157.766  &  -48.203  &  7.01$\, \pm \,  $0.08  &  7.45$\, \pm \,  $0.08  &  1.06$\, \pm \,  $0.02  &  0.18$\, \pm \,  $0.08  \\ 
3C75-1  &  HT03  &  170.216  &  -44.911  &  8.63$\, \pm \,  $0.30  &  9.21$\, \pm \,  $0.30  &  1.07$\, \pm \,  $0.05  &  0.23$\, \pm \,  $0.09  \\ 
3C75-2  &  HT03  &  170.296  &  -44.919  &  8.63$\, \pm \,  $0.30  &  9.20$\, \pm \,  $0.30  &  1.07$\, \pm \,  $0.05  &  0.24$\, \pm \,  $0.09  \\ 
3C79  &  HT03  &  164.149  &  -34.457  &  9.46$\, \pm \,  $0.23  &  10.20$\, \pm \,  $0.25  &  1.08$\, \pm \,  $0.03  &  0.30$\, \pm \,  $0.11  \\ 
3C98-1  &  HT03  &  179.859  &  -31.086  &  10.34$\, \pm \,  $0.19  &  11.48$\, \pm \,  $0.22  &  1.11$\, \pm \,  $0.03  &  0.22$\, \pm \,  $0.08  \\ 
3C98-2  &  HT03  &  179.829  &  -31.024  &  10.45$\, \pm \,  $0.19  &  11.68$\, \pm \,  $0.23  &  1.12$\, \pm \,  $0.03  &  0.20$\, \pm \,  $0.08  \\ 
3C98  &  HT03  &  179.837  &  -31.049  &  10.40$\, \pm \,  $0.22  &  11.94$\, \pm \,  $0.27  &  1.15$\, \pm \,  $0.03  &  0.29$\, \pm \,  $0.11  \\ 
4C07.32  &  HT03  &  322.228  &  68.835  &  2.63$\, \pm \,  $0.07  &  2.69$\, \pm \,  $0.07  &  1.02$\, \pm \,  $0.04  &  0.12$\, \pm \,  $0.06  \\ 
4C13.65  &  HT03  &  39.315  &  17.718  &  9.86$\, \pm \,  $0.39  &  10.50$\, \pm \,  $0.41  &  1.06$\, \pm \,  $0.06  &  0.22$\, \pm \,  $0.08  \\ 
4C20.33  &  HT03  &  20.185  &  66.834  &  2.26$\, \pm \,  $0.10  &  2.28$\, \pm \,  $0.10  &  1.01$\, \pm \,  $0.07  &  0.18$\, \pm \,  $0.07  \\ 
P0320+05  &  HT03  &  176.982  &  -40.843  &  10.73$\, \pm \,  $0.31  &  12.02$\, \pm \,  $0.37  &  1.12$\, \pm \,  $0.04  &  0.40$\, \pm \,  $0.15  \\ 
P0347+05  &  HT03  &  182.274  &  -35.731  &  12.76$\, \pm \,  $0.25  &  15.34$\, \pm \,  $0.32  &  1.20$\, \pm \,  $0.03  &  0.28$\, \pm \,  $0.10  \\ 
P0428+20  &  HT03  &  176.808  &  -18.557  &  19.30$\, \pm \,  $0.62  &  28.01$\, \pm \,  $0.80  &  1.45$\, \pm \,  $0.04  &  0.43$\, \pm \,  $0.16  \\ 
P0820+22  &  HT03  &  201.364  &  29.676  &  4.86$\, \pm \,  $0.14  &  4.88$\, \pm \,  $0.14  &  1.00$\, \pm \,  $0.04  &  0.08$\, \pm \,  $0.04  \\ 
P1055+20  &  HT03  &  222.510  &  63.130  &  1.66$\, \pm \,  $0.07  &  1.67$\, \pm \,  $0.08  &  1.01$\, \pm \,  $0.06  &  0.08$\, \pm \,  $0.04  \\ 
P1117+14  &  HT03  &  240.438  &  65.788  &  2.39$\, \pm \,  $0.17  &  2.39$\, \pm \,  $0.17  &  1.00$\, \pm \,  $0.10  &  0.00$\, \pm \,  $0.01  \\ 
\enddata
\tablecomments{ (1): Target name; (2): Survey, either MACH (this work), 21-SPONGE \citep{murray2018b}, Perseus \citep{stanimirovic2014}, HT03 \citep{heiles2003}. (3,4): Galactic longitude and latitude ($l$, $b$) coordinates; (5): \hi\ column density in the optically-thin limit; (6): Total \hi\ column density (Equation~\ref{e:nhi}); (7): Column density correction for optical depth (\rhi; Equation~\ref{e:rhi}); (8): Fraction of CNM (\fcnm; Equation~\ref{e:fcnm}). } 
\end{deluxetable*}

\section{Comparison with SPONGE Fits}
\label{ap:sponge_comparison}

\begin{figure*}
\begin{center}
\includegraphics[width=\textwidth]{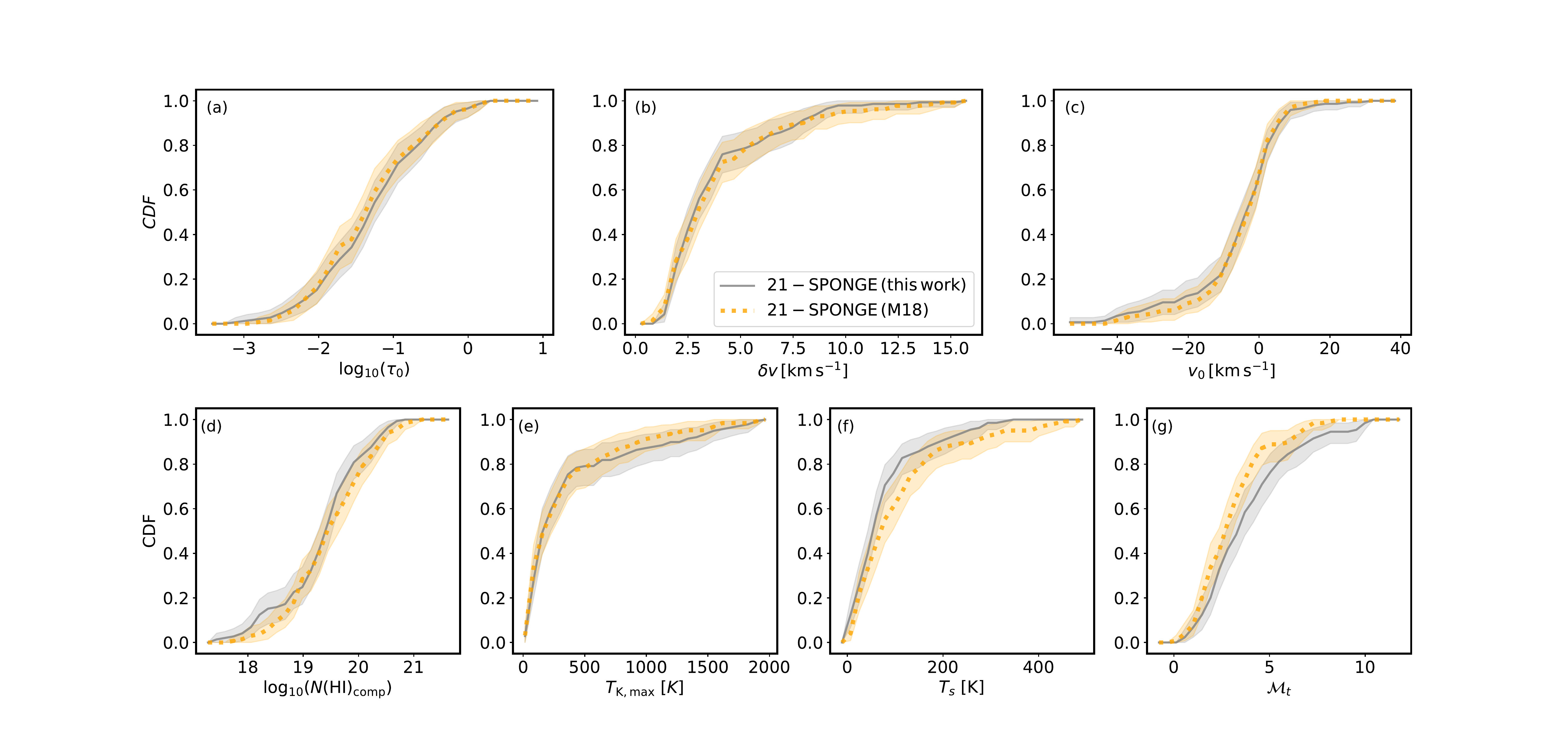}
\caption{Comparison of cumulative distribution functions (CDFs) of parameters from the Gaussian fits to 21-SPONGE sources from this work (grey) and the analysis of \citet[][; yellow (M18)]{murray2018b}, including amplitude ($\tau_{0}$; a), line width (FWHM; $\delta v$; b), mean velocity ($v_{0}$; c), and derived physical properties including column density (Equation~\ref{e:nhi_comp}, $N({\rm HI})_{\rm comp}$; e), maximum kinetic temperature (Equation~\ref{e:tkmax_comp}, $T_{\rm K, max}$; e), spin temperature ($T_s$; f), and turbulent Mach number (Equation~\ref{e:mt_comp}, $\mathcal{M}_t$; g). The uncertainty ranges in the CDFs are computed  by bootstrapping the sample with replacement over $10^4$ trials, and illustrate the $1^{st}$ through $99^{\rm th}$ percentiles. We find that the method presented in this work (solid) is fully consistent with the results of the original 21-SPONGE analysis.}
\label{f:sponge_comparison}
\end{center}
\end{figure*}

In the following appendix, we present the results of a comparison between the \hi\ decomposition method used in this work (Section~\ref{sec:gauss_fits}) with the results of \citet{murray2018b} for the 21-SPONGE sample. For overlapping targets (i.e., all 21-SPONGE sources with $|b|>10^{\circ}$) we compare the fitted Gaussian parameters for individual \hi\ structures ($\tau_0$, $\delta v$, $v_0$), and their inferred physical properties (column density ($N({\rm HI})_{\rm comp}$), maximum kinetic temperature ($T_{k,\rm max}$), spin temperature ($T_s$) and turbulent MACH number ($M_t$)). We plot cumulative distribution functions of these properties in Figure~\ref{f:sponge_comparison}. Within uncertainties, estimated via bootstrapping each sample with replacement, the distributions of all properties are the same. This gives us confidence that our new  method, which involves using $21\rm\,cm$ emission from EBHIS across \noff\ off-positions to infer the fitted parameters, is fully consistent with previous work. 

\section{Gaussian Fits to the Comparison Sample}
\label{ap:comparison_fits}

\begin{figure*}
\begin{center}
\vspace{-80pt}
\includegraphics[width=\textwidth]{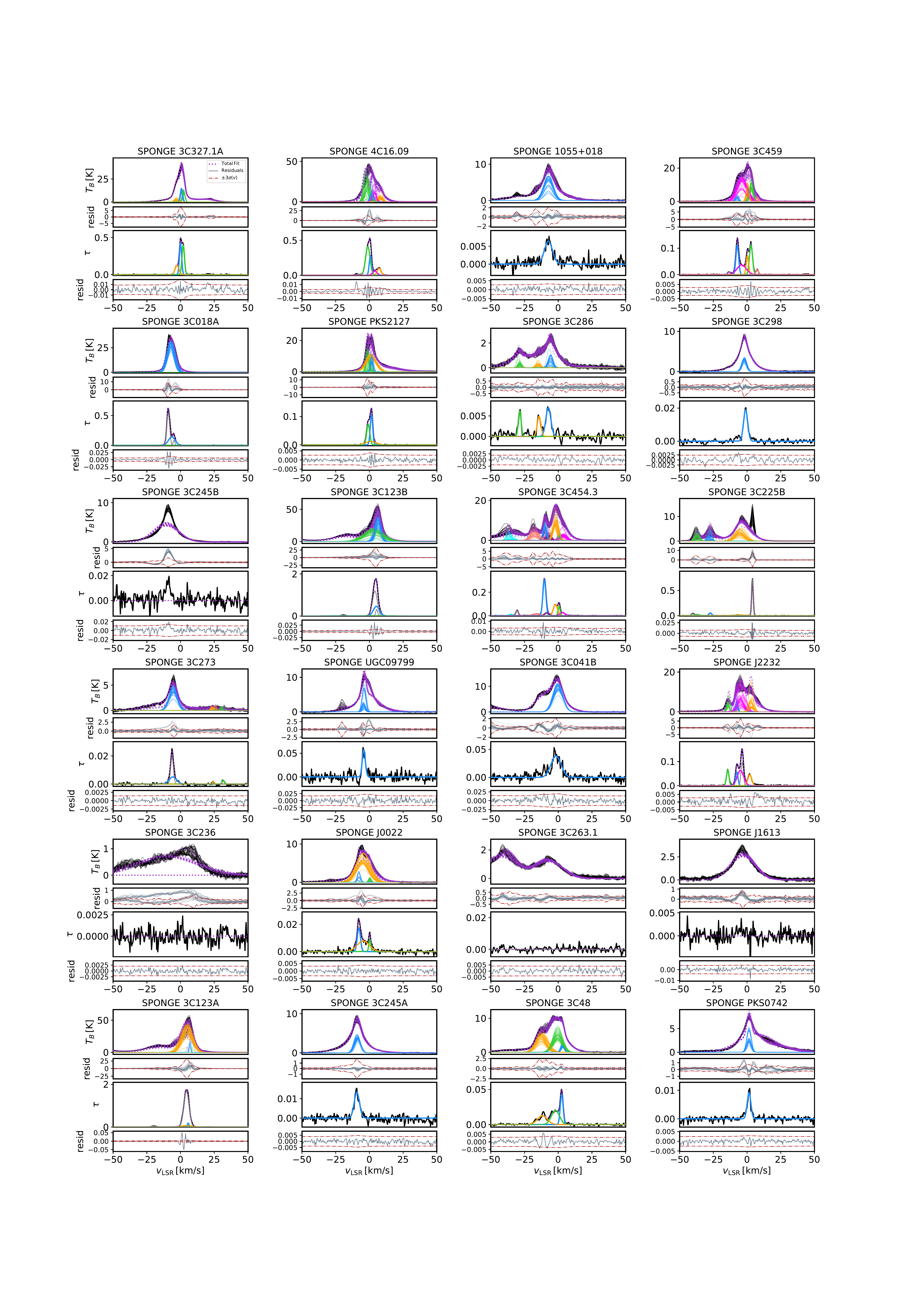}
\vspace{-80pt}
\caption{Gaussian fits to the comparison sample (SPONGE, HT03 and Perseus) \thi\ and $T_B(v)$ pairs using GaussPy. Each panel corresponds to a different sightline, including $T_B(v)$ from the \noff\ off-positions (top panel), and \thi\ (bottom panel), with the total fits (dotted purple) to \thi\ \citep[Equation 1;][]{murray2018b} and $T_B(v)$ (Equation~\ref{e:tbfit}), the individual components (solid, colors) and the residuals from the fits (sub-panels; grey), along with $\pm3\sigma$ uncertainties (red).}
\label{f:comparison_decomposition}
\end{center}
\end{figure*}

In the following Appendix we present the results of autonomous, simultaneous decomposition of the comparison sample (21-SPONGE, Perseus, HT03). Figure~\ref{f:comparison_decomposition} includes panels showing $T_B(v)$ (top panel) and \thi\ (bottom panel) for each of the \nrest\ LOS, including the best-fit components in absorption and their corresponding emission.

\begin{figure*}
\begin{center}
% \ContinuedFloat
\vspace{-80pt}
\includegraphics[width=\textwidth]{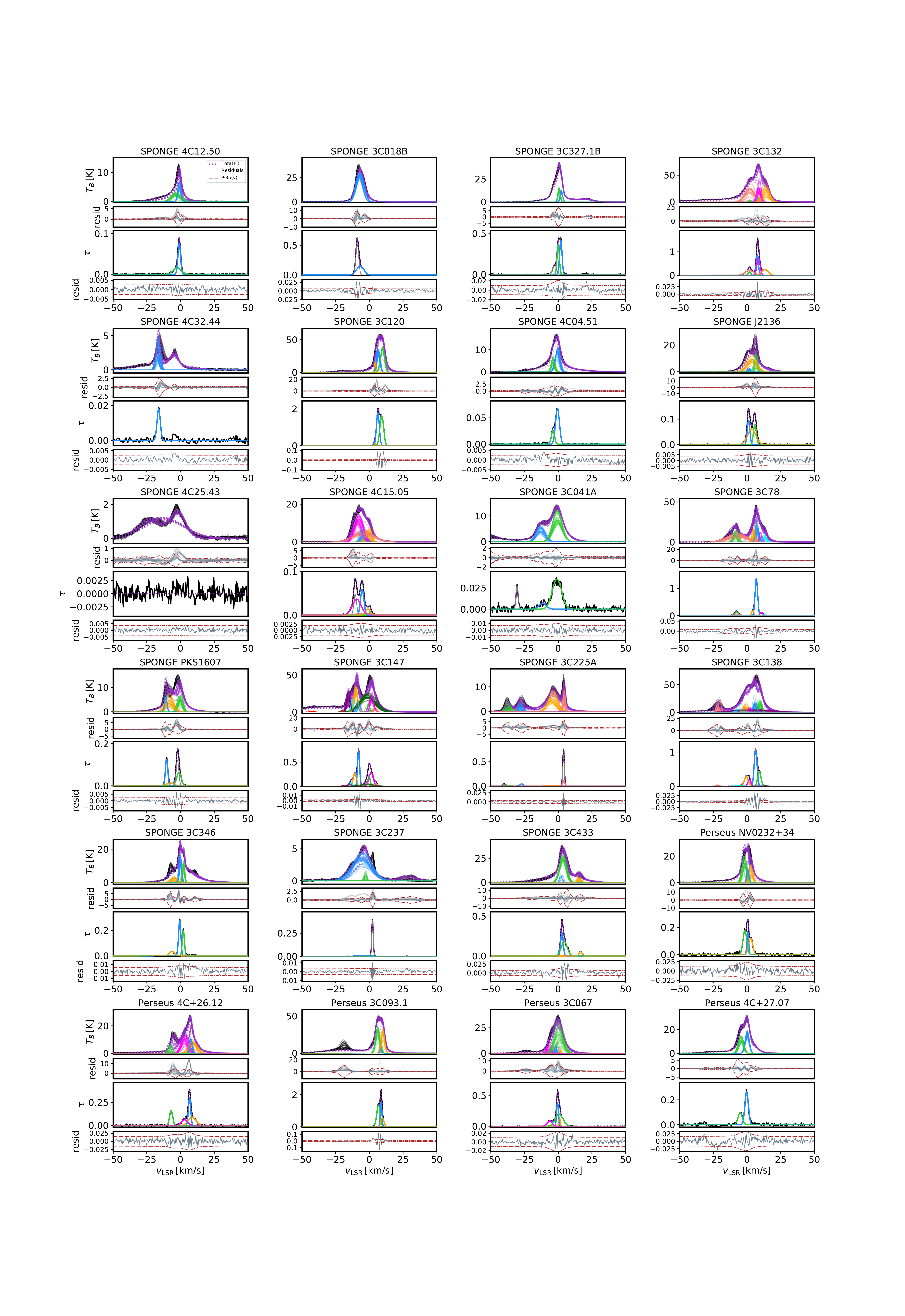}
% \vspace{-80pt}
% \caption{(contd)}
% \label{f:comparison_decomposition_b}
\end{center}
\end{figure*}

\begin{figure*}
\begin{center}
% \ContinuedFloat
\vspace{-80pt}
\includegraphics[width=\textwidth]{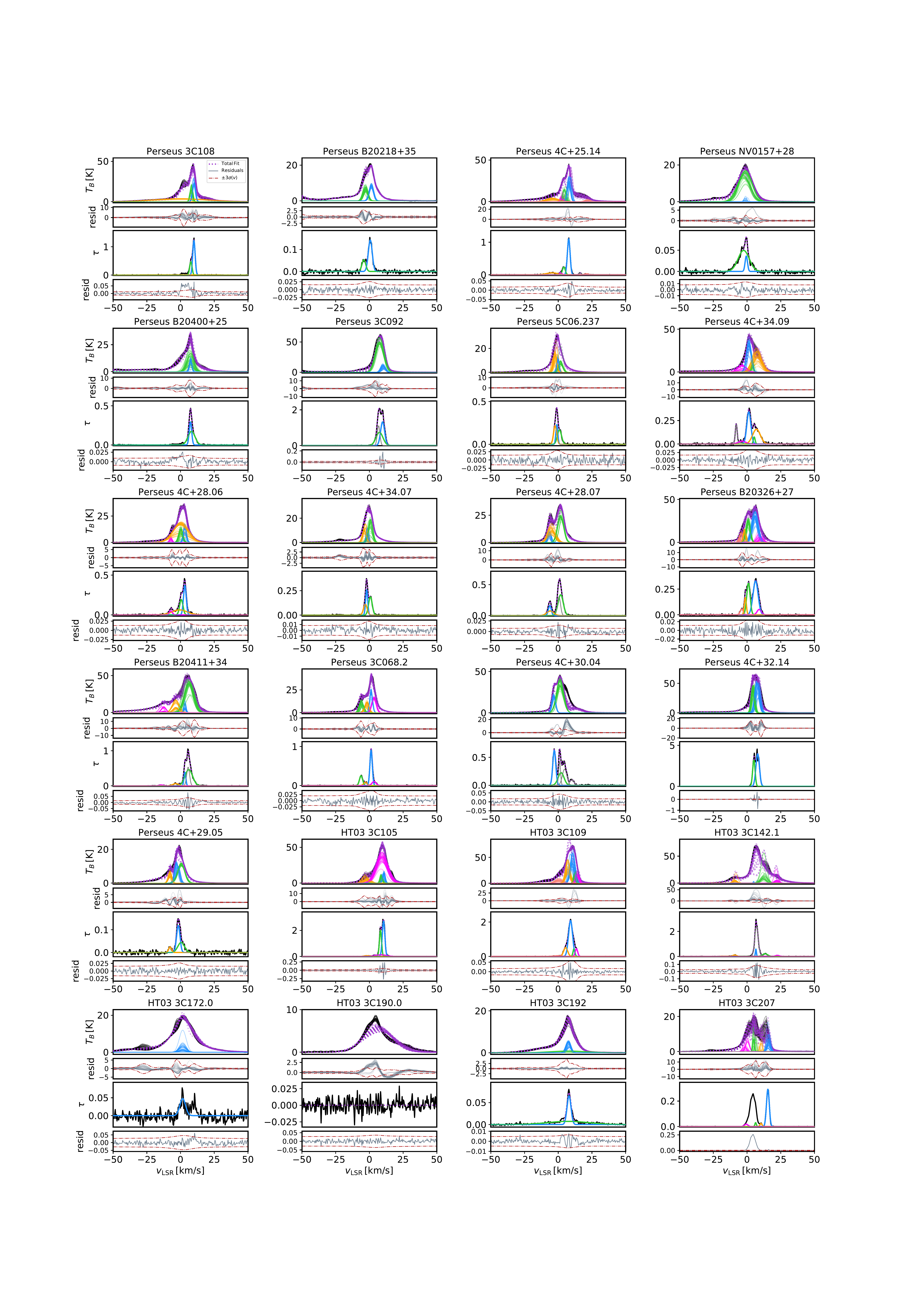}
% \vspace{-80pt}
% \caption{(contd)}
% \label{f:comparison_decomposition_c}
\end{center}
\end{figure*}

\begin{figure*}
\begin{center}
% \ContinuedFloat
\vspace{-80pt}
\includegraphics[width=\textwidth]{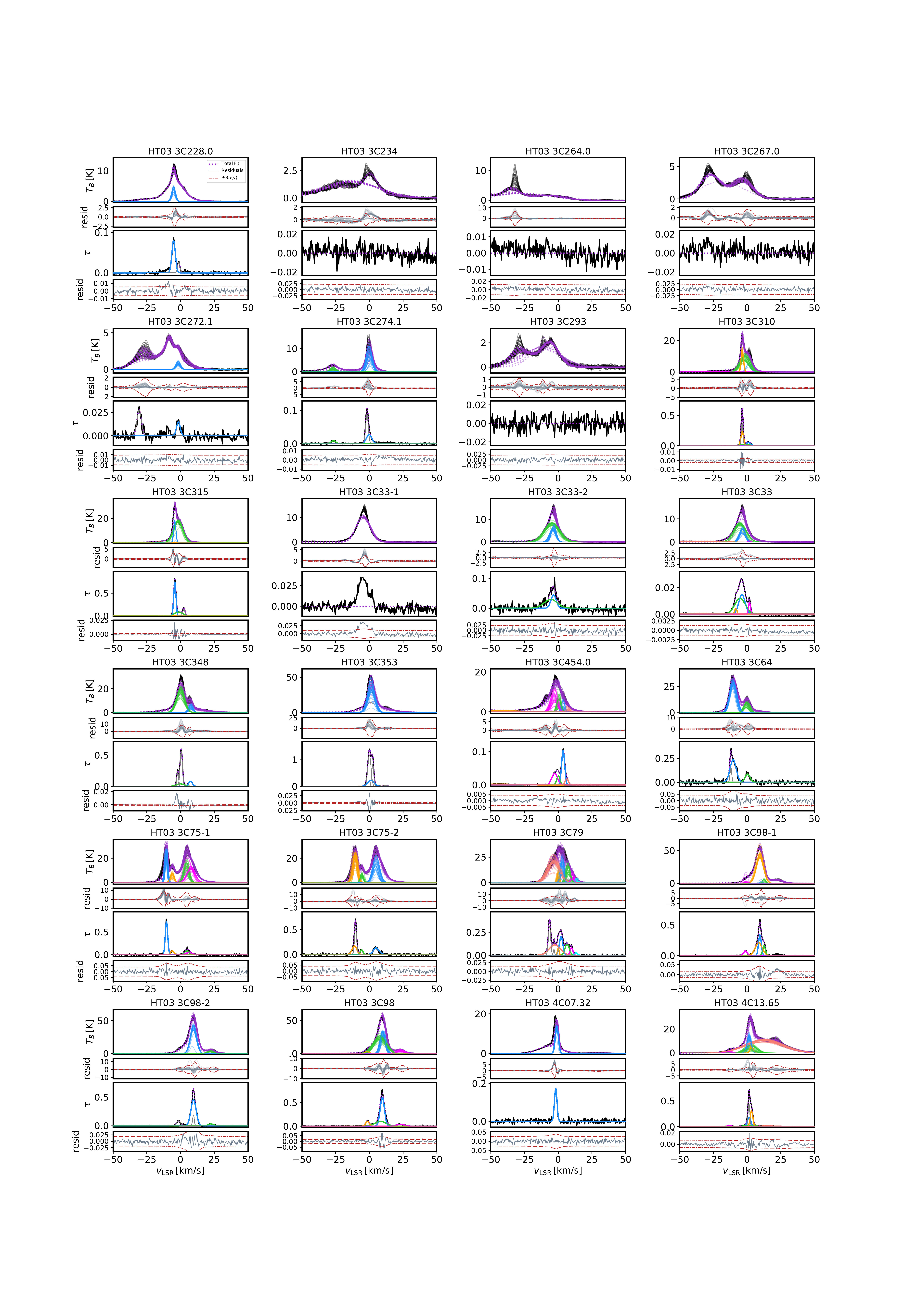}
% \vspace{-80pt}
% \caption{(contd)}
% \label{f:comparison_decomposition_d}
\end{center}
\end{figure*}

\begin{figure*}
\begin{center}
% \ContinuedFloat
\vspace{-80pt}
\includegraphics[width=\textwidth]{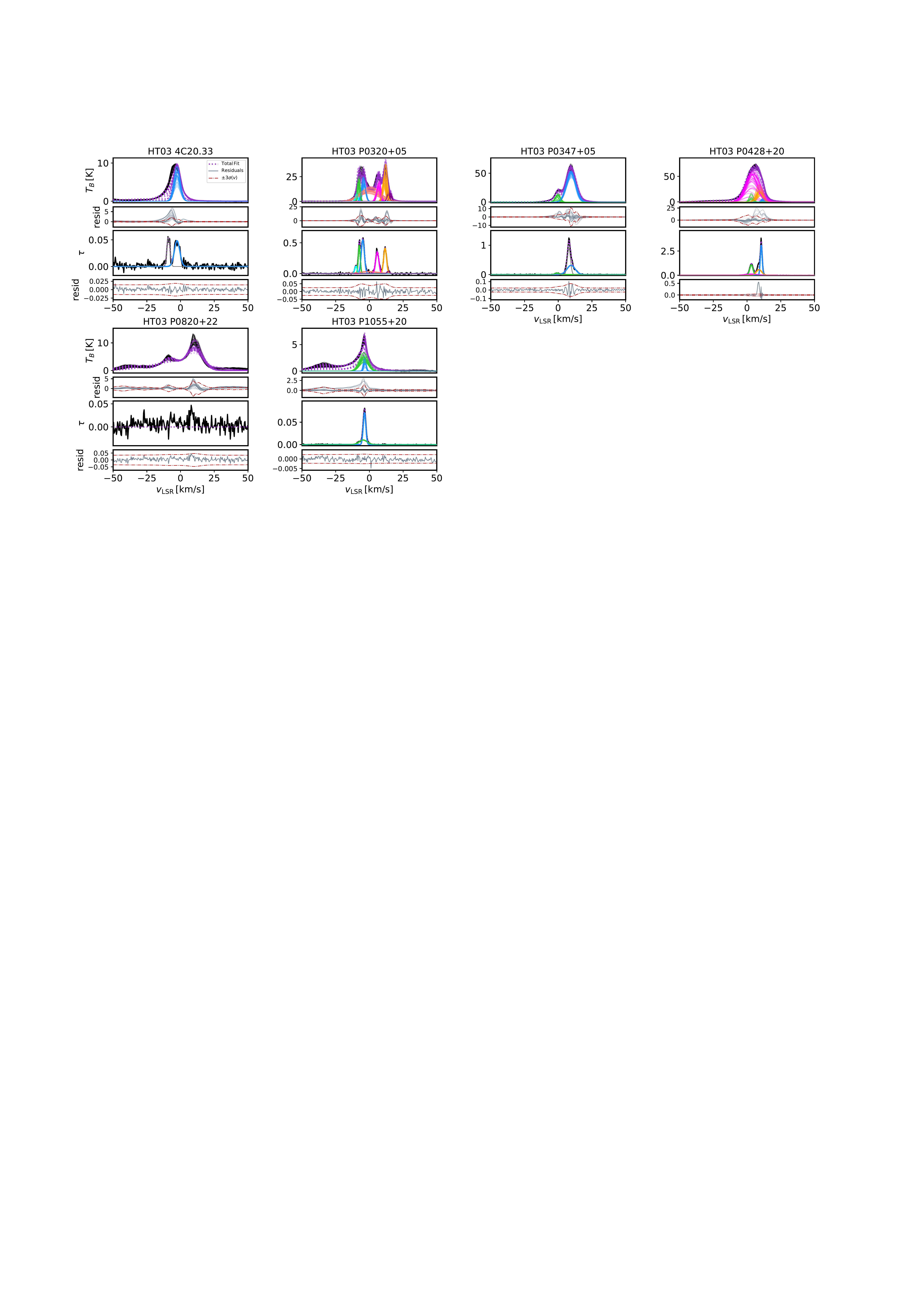}
% \vspace{-470pt}
% \caption{(contd)}
% \label{f:comparison_decomposition_e}
\end{center}
\end{figure*}

\bibliography{ms}
\bibliographystyle{aasjournal}

\end{document}